\documentclass{IEEEtran}

\usepackage{amsmath,amssymb,amsfonts,mathtools}
\usepackage{bm}
\usepackage{booktabs}
\usepackage{array}
\usepackage{algorithm}
\usepackage{algorithmic}
\usepackage{cite}
\usepackage{graphicx}
\usepackage{url}
\usepackage{multirow}
\usepackage{enumitem}

\usepackage{mathrsfs}
\usepackage{amsthm,amsmath,amssymb,amsfonts,mathtools}
\usepackage{epstopdf}
\usepackage{flushend}
\usepackage{colortbl}
\usepackage[table]{xcolor}
\usepackage{tabularx}
\usepackage{makecell}
\usepackage{arydshln}

\usepackage{float}
\usepackage{needspace}
\allowdisplaybreaks[4]

\newcolumntype{Y}{>{\centering\arraybackslash}X}
\newcommand{\R}{\mathbb{R}}
\newcommand{\Ssym}{\mathbb{S}}
\newcommand{\col}{\operatorname{col}}

\newcommand{\ip}[2]{\left\langle #1,#2\right\rangle}

\newtheorem{definition}{Definition}

\newtheorem{remark}{Remark}
\newtheorem{theorem}{Theorem}

\newtheorem{proposition}{Proposition}
\newtheorem{corollary}{Corollary}

\usepackage{comment}
\newboolean{showcomments}
\setboolean{showcomments}{false}

\newcommand{\fliu}[1]{\ifthenelse{\boolean{showcomments}}
        { \textcolor{red}{(FL:  #1)}}{}}
\newcommand{\xru}[1]{\ifthenelse{\boolean{showcomments}}
        { \textcolor{blue}{(XR:  #1)}}{}}

\title{Port-Shaping Dissipativity for Compositional Stability Analysis of Heterogeneous Power Systems
\thanks{This paper is supported by the Smart Grid National Science and Technology Major Project under Grant 2024ZD0801200 and the Science and Technology Project of China Southern Power Grid Co., Ltd under Grant GDKJXM20250201(036000KC25020005). (Corresponding author: Feng Liu.)}
\thanks{Xi Ru, Xiaoyu Peng and Feng Liu are with the Department of Electrical Engineering, Tsinghua University, Beijing 100084, China (e-mail: thu.ruxi@gmail.com; pengxy19@tsinghua.org.cn; lfeng@tsinghua.edu.cn)}
\thanks{Jinchang Chen is with Power Dispatch Control Center, Guangdong Power Grid Corporation, Guangzhou 510530, China (e-mail: chenjinchang\_2010@163.com)}
}

\author{\vspace{-2\baselineskip}%
Xi Ru,~\IEEEmembership{ Student Member,~IEEE}, Xiaoyu Peng,~\IEEEmembership{Graduate Student Member,~IEEE}, Jinchang Chen, and Feng Liu,~\IEEEmembership{Senior Member,~IEEE}%
}

\begin{document}
\maketitle

\begin{abstract}
Dissipativity-based compositional stability analysis is attractive for large-scale interconnected systems. However, its application to complex heterogeneous power systems is limited by conservative nodal dissipativity descriptions and inefficient distributed implementation. In this paper, we introduce a unified dynamic port-shaping framework through a common first-order augmentation of the physical voltage–current ports, enabling complementary frequency-dependent supply descriptions while preserving the physical feedback interconnection structure through the network. The resulting representation accommodates heterogeneous nonlinear power system bus dynamics within a unified dissipativity description and enables more accurate compositional stability analysis. We then formulate and solve a centralized stability-certification problem through a dual-guided column-generation scheme. We show that its matrix dual identifies the network-critical dissipation direction, guiding parallel local certificate generation toward the current system bottleneck. This leads to a distributed Network-Critical Margin Exchange (NCME) method with early stability certification. Case studies demonstrate substantially reduced conservatism through dynamic port shaping and improved computational scalability of NCME for heterogeneous power systems.
\end{abstract}


\section{Introduction}

Power system stability analysis is fundamental to the secure operation of interconnected grids~\cite{Kundur1994,Hatziargyriou2021}. With the increasing penetration of renewable generation and power-electronic interfaces, modern power systems are becoming larger, more heterogeneous, and dynamically more complex~\cite{Hatziargyriou2021,Tayyebi2020}. Synchronous machines, grid-forming and grid-following converters, dynamic and static loads, and diverse control architectures exhibit substantially different dynamics, operating time scales, and terminal causalities. These developments call for stability-analysis methods that can accommodate heterogeneous nodal behavior while remaining scalable to large networks.

Dissipativity provides a natural framework for this purpose because subsystem properties can be characterized locally through storage functions and supply rates. At the same time, system-level stability follows from compatibility of the local dissipative properties under interconnection~\cite{Willems1972a,Willems1972b,ArcakBook}. Another advantage of dissipativity is its applicability to nonlinear dynamical systems. Depending on the system structure, nonlinear dissipativity conditions can be formulated through Hamilton–Jacobi inequalities~\cite{van2000l2}, computationally certified using sum-of-squares (SOS) programming for polynomial systems~\cite{Parrilo2000SOS, PapachristodoulouPrajna2005SOS}, or combined with multiplier and integral quadratic constraint (IQC) methods~\cite{MegretskiRantzer1997} to characterize nonlinearities. As a specialization of dissipativity, passivity is particularly attractive for distributed analysis and has been widely used in power systems~\cite{Khodabakhshloo2026,Harnefors2016,Nahata2020}. However, a prescribed passivity relation can be restrictive because heterogeneous devices find it difficult to be passive over all frequency ranges under the same input--output ports~\cite{DeyKulkarni2021,Laib2022}. Generalized formulations, including differential passivity~\cite{Yang2020,Li2026GFMDifferential,li2024distributed}, matrix-valued indices~\cite{ru2026conservativedistributedstabilityanalysis}, and equilibrium-free dissipativity~\cite{Yang2025PartI,Yang2025PartII}, extend the applicability, but these methods still rely on device-dependent supply structures. Depending on the formulation, this may restrict the applicable device classes or frequency ranges, or require network reductions to eliminate inapplicable buses. Consequently, existing descriptions can be difficult to combine across heterogeneous buses without restrictions or loss of dissipativity information, limiting their applicability to compositional stability analysis of power systems.

Dynamic supply rates and related integral quadratic constraint (IQC) formulations provide general tools for representing frequency-dependent dissipativity through auxiliary dynamic operators~\cite{ArcakBook,TippettBao2014,Seiler2015,KhongChenLanzon2025}. Recent power-system studies have also explored these approaches. Dynamic frequency-dependent passivity multipliers have been introduced to enlarge plug-and-play small-signal stability certificates for converter-dominated grids~\cite{gorbunov2027dynamic}. At the same time, terminal-behavior contracts combined with IQC-based and structured robust-stability conditions have been developed for compositional stability assessment of multi-vendor converter-dominated systems~\cite{qin2026minimal}.
These studies demonstrate the value of frequency-dependent dissipativity descriptions in power-system stability analysis. However, a unified dynamic supply representation that preserves the physical voltage–current port structure and original network projection while supporting nonlinear time-domain dissipativity certification remains lacking. In particular, jointly optimizing local supply matrices for heterogeneous dynamic and algebraic buses under a common network compatibility condition remains challenging.

Exploiting such complementary dissipativity further requires coordinating the local supply matrices through a network-level matrix inequality, leading to a coupled semidefinite program. Existing compositional-certification methods address this coupling through dual decomposition or distributed optimization, including alternating direction method of multipliers (ADMM)-based approaches~\cite{Yang2025PartII,Topcu2009,Meissen2015,BoydADMM2011}. While systematic, these methods do not directly exploit network-critical dissipation directions to guide local certificate searches, leaving opportunities to improve computational and communication efficiency in large-scale heterogeneous power systems.

To address these challenges, this paper develops a unified dynamic port-shaping framework for heterogeneous power-system buses through a common first-order augmentation of the physical voltage–current port. The resulting dynamic supply rate captures complementary frequency-dependent terminal behavior while retaining a common network composition map. Building on this representation, the compositional stability-certification problem is further decomposed through dual-guided column generation, where the matrix dual reveals the network-critical dissipation direction and coordinates parallel local certificate searches. The main contributions are summarized as follows.
\begin{itemize}[leftmargin=1em,labelsep=0.4em,itemindent=0pt,listparindent=0pt]
\item \textbf{Unifying dissipativity descriptions of heterogeneous power-system buses via dynamic port shaping}. A common first-order dynamic shaping of the voltage–current port establishes a complementary frequency decomposition while preserving the original network interconnection structure. This enriched representation unifies diverse nodal dissipativity descriptions within a common dynamic supply-rate framework, enabling more capable dissipativity matching, reducing conservatism, and extending the applicability of compositional power-system stability analysis.

\item \textbf{Accelerating distributed stability certification using the critical dissipation direction of networks}. We formulate the centralized compositional certification problem and solve it via column generation. We show that the matrix dual identifies the network-critical dissipation direction and translates the system-level bottleneck into informative local weightings for certificate generation. This physics-informed coordination enables distributed stability certification with early termination, improving the computational scalability and reducing the coordination burden of distributed stability analysis for large heterogeneous power systems.
\end{itemize}

The remainder of this paper is organized as follows. Section II introduces the preliminaries and power-system model. Section III develops the unified dynamic dissipativity port and the compositional stability criterion. Section IV formulates convex local dissipativity certificates and the centralized stability-verification problem. Section V presents the dual interpretation and the distributed NCME method. Section VI reports case studies, and Section VII concludes the paper.

\textit{Notation:}
$\mathbb{R}^n$ and $\mathbb{S}^n$ denote the $n$-dimensional real vector space and the set of real symmetric $n\times n$ matrices, respectively. For a vector of indeterminates \(x\), \(\mathbb R[x]\) denotes the ring of real polynomials in \(x\), and \(\mathbb R[x]^m\) denotes the space of \(m\)-dimensional vectors of such polynomials. For $A,B\in\mathbb{S}^n$, $A\succeq B$ ($A\succ B$) means that $A-B$ is positive semidefinite (positive definite), $A^H$ means the Hermitian transpose of $A$. $I_n$ and $0_{m\times n}$ denote the $n$-dimensional identity matrix and zero matrix, respectively, and $e_i$ is the $i$th canonical basis vector. The operators $\operatorname{col}(\cdot)$, $\operatorname{diag}(\cdot)$, and $\otimes$ denote column stacking, block-diagonal concatenation, and the Kronecker product. For matrices $A$ and $B$ of compatible dimensions, $\langle A,B\rangle:=\operatorname{tr}(A^\top B)$ denotes the Frobenius inner product, while $\ker A$ and $\operatorname{range}(A)$ denote the nullspace and range of $A$. For $A=A^\top$ and $B\succ0$, $\lambda_{\max}(A,B)
:=\max_{x\neq0}\frac{x^\top Ax}{x^\top Bx}
=\lambda_{\max}(B^{-1/2}AB^{-1/2})$
denotes the largest generalized eigenvalue of $(A,B)$.
For a set \(\mathcal S\), \(\operatorname{conv}\mathcal S\) denotes its convex hull. $\widehat a(s)=\mathcal L\{a(t)\}(s)$
denotes the Laplace transform of a time-domain signal $a(t)$.

\section{Modeling and Preliminaries}

\subsection{Dissipativity and Passivity}
Consider a dynamical system:
\begin{equation}\label{eq-dynamic}
        \dot{x}=f(x,u),\quad
        y=h(x,u)
\end{equation}
where state $x\in\mathcal{X}\subset\mathbb{R}^n$, input $u\in \mathcal{U}\subset\mathbb{R}^m$, and output $y\in\mathcal{Y}\subset\mathbb{R}^m$. The map $f:\mathcal{X}\times \mathcal{U}\rightarrow\mathbb{R}^n$ is locally Lipschitz, and $h:\mathcal{X}\times \mathcal{U}\rightarrow\mathcal{Y}$ is continuous. Without loss of generality, we assume the origin is an equilibrium of~\eqref{eq-dynamic}. That is, $f(0,0)=0$, $h(0,0)=0$. Denote as $z(t)=\mathrm{col}(u(t),y(t))$. The definition of dissipativity is shown below.

\begin{definition}[Dissipativity\cite{van2000l2}]
    The system~\eqref{eq-dynamic} is dissipative with respect to a supply rate $w(z(t))$ if there exists a positive semidefinite (p.s.d.) function $V(x)$, referred to as the storage function, s.t. for all initial conditions $x(t_0)\in\mathcal{X}$, all admissible inputs $u$ and all $t_1\ge t_0$, $V(x(t_1))\le V(x(t_0))+\int_{t_0}^{t_1}w(z(t))\, \mathrm{d} t$.
    \label{def-dissipativity}
\end{definition}

For a continuously differentiable storage function, Definition~\ref{def-dissipativity} shows that system~\eqref{eq-dynamic} is dissipative with respect to a supply rate $w(z)$ if every admissible trajectory satisfies $\dot V(x(t)) \le w(z(t))$. A widely used class of supply rates is the quadratic form $w(z)=z^\top X z$, $X=X^\top$, also called QSR-dissipativity, where $X$ is called the supply matrix. Passivity and strict passivity~\cite{khalil2002nonlinear} are particular choices of $X$, that is 
$X=\begin{bmatrix}
        -\rho I ,\, \frac{1}{2}I;\,\frac{1}{2}I ,\, -\mu I
    \end{bmatrix}$, $\quad \rho,\mu\ge 0$.
Differential-passivity~\cite{Yang2020,ru2026conservativedistributedstabilityanalysis,li2025passivity} formulations extend this idea with derivative information. Representative output-differential passivity (ODP) and input-differential passivity (IDP) structures can be written schematically as $\dot V\le u^\top \dot y -\rho y^\top\dot y,\rho\ge 0$ and $\dot V\le\dot u^\top  y -\nu \dot u^\top u, \nu\ge 0$, respectively. 

\subsection{Structure-Preserving DAE modeling of power systems}
Consider a structure-preserving power system composed of $N$ buses and transmission lines connecting them. It can be represented by an undirected graph
$\mathcal{G}=(\mathcal{N},\mathcal{E})$, where $\mathcal{N}$ and $\mathcal{E}$
denote the sets of buses and transmission lines, respectively. 
At bus $i$, define the terminal voltage and current as $\mathbf{v}_i =
    \begin{bmatrix}
        V_{D,i},\,
        V_{Q,i}
    \end{bmatrix}^\top$, $\mathbf{i}_i =
    \begin{bmatrix}
        I_{D,i},\,
        I_{Q,i}
    \end{bmatrix}^\top$ where the subscript $D$ and $Q$ denotes the $D$-axis and $Q$-axis components of the port voltage $V_i$ or port current $I_i$, respectively. $\mathbf{i}_i$ is defined positive from the network into the local bus subsystem.

\begin{table*}[!t]
\centering
\caption{Representative heterogeneous power-system nodes and their conventional dissipativity descriptions}
\label{tab:heterogeneous_nodes}
\renewcommand{\arraystretch}{1.15}
\setlength{\tabcolsep}{4pt}
\begin{tabular}{
>{\centering\arraybackslash}m{4.1cm}
>{\centering\arraybackslash}m{3.3cm}
>{\centering\arraybackslash}m{3.0cm}
>{\centering\arraybackslash}m{3.2cm}
>{\centering\arraybackslash}m{3.1cm}}
\toprule
\textbf{Device class}
& \textbf{Temporal characteristics}
& \textbf{Dissipativity description}
& \textbf{Dissipativity coordinates}
& \textbf{Representative port causality} \\
\midrule

Synchronous generator (SG)
\cite{Yang2020,DeyKulkarni2021}
& Electromechanical dynamics
& ODP/passivity-type
& Power/voltage-angle variables
& $v_i\rightarrow i_i$ \\

Grid-forming converter with virtual synchronous generator (VSG) control
\cite{Yang2025PartII,Li2026GFMDifferential}
& Outer-loop dynamics
& ODP-type
& Power/voltage-angle variables
& $i_i\rightarrow v_i$ \\

Grid-forming converter with conventional droop (CD) control
\cite{Yang2025PartII,HeDorfler2024,Li2026GFMDifferential}
& Outer-loop dynamics
& ODP-type
& Power/voltage-angle variables
& $i_i\rightarrow v_i$ \\

Grid-forming converter with quadratic droop (QD) control
\cite{SimpsonPorco2017QuadraticDroop,Yang2020,ru2026conservativedistributedstabilityanalysis}
& Outer-loop dynamics
& ODP-type
& Power/voltage-angle variables
& $i_i\rightarrow v_i$ \\

PLL-synchronized grid-following converter (GFL)
\cite{Yang2025PartII,li2025passivity}
& Synchronization and power-control dynamics
& IDP-type
& Power/voltage-angle variables
& $v_i\rightarrow i_i$ \\

ZIP load
\cite{Nahata2020,Yang2025PartII}
& Static
& Static QSR
& Voltage--current variables
& $v_i\rightarrow i_i$ \\

Constant-impedance load
\cite{Nahata2020,Yang2025PartII}
& Static
& Static QSR
& Voltage--current variables
& $v_i\rightarrow i_i$ \\

Intermediate / junction bus
\cite{Yang2025PartII}
& Static
& Static QSR
& Voltage--current variables
& $v_i\rightarrow i_i$ \\

Constant-voltage source
\cite{Yang2025PartII}
& Static
& Static QSR
& Voltage--current variables
& $i_i\rightarrow v_i$ \\

\bottomrule
\end{tabular}
\end{table*}

The bus set is partitioned as
$\mathcal{N}=\mathcal{N}_{\rm d}\cup\mathcal{N}_{\rm s}$, $\mathcal{N}_{\rm d}\cap\mathcal{N}_{\rm s}=\varnothing$, where $\mathcal{N}_{\rm d}$ contains buses with internal dynamic states and $\mathcal{N}_{\rm s}$ contains static or purely algebraic buses. For each $i\in\mathcal{N}_{\rm d}$, the local dynamics are described in the general nonlinear form $\Sigma_i:
    \dot{x}_i = f_i(x_i,u_i),\quad
    y_i= h_i(x_i,u_i)$
where $x_i\in\mathbb{R}^{n_i}$ is the internal state, $n_i$ is the dimension of the dynamic state of bus $i$, and $u_i,y_i\in\mathbb{R}^{2}$ are terminal inputs and outputs.
For each $i\in\mathcal{N}_{\rm s}$, the corresponding static relation is written as $\Sigma_i: y_i = h_i(u_i)$.
These models encompass substantially different dynamic and dissipative characteristics, as summarized in Table \ref{tab:heterogeneous_nodes}. For example, synchronous generators may exhibit low-frequency nonpassivity but become passive at higher frequencies at the physical voltage–current port~\cite{DeyKulkarni2021}, motivating output-differential or negative-imaginary descriptions~\cite{Yang2020,Khodabakhshloo2026}. In contrast, static network elements and loads have no internal dynamics and are naturally characterized by static quadratic voltage–current relations. Port causality also depends on device type and control architecture. PLL-based grid-following converters typically admit a voltage-input/current-output description, whereas grid-forming converters are more naturally represented with the opposite orientation~\cite{li2025passivity,Yang2025PartII}. Detailed heterogeneous device models and their dissipativity derivations are provided in Appendix A.

\begin{figure}
    \centering
    \includegraphics[width=1.0\linewidth]{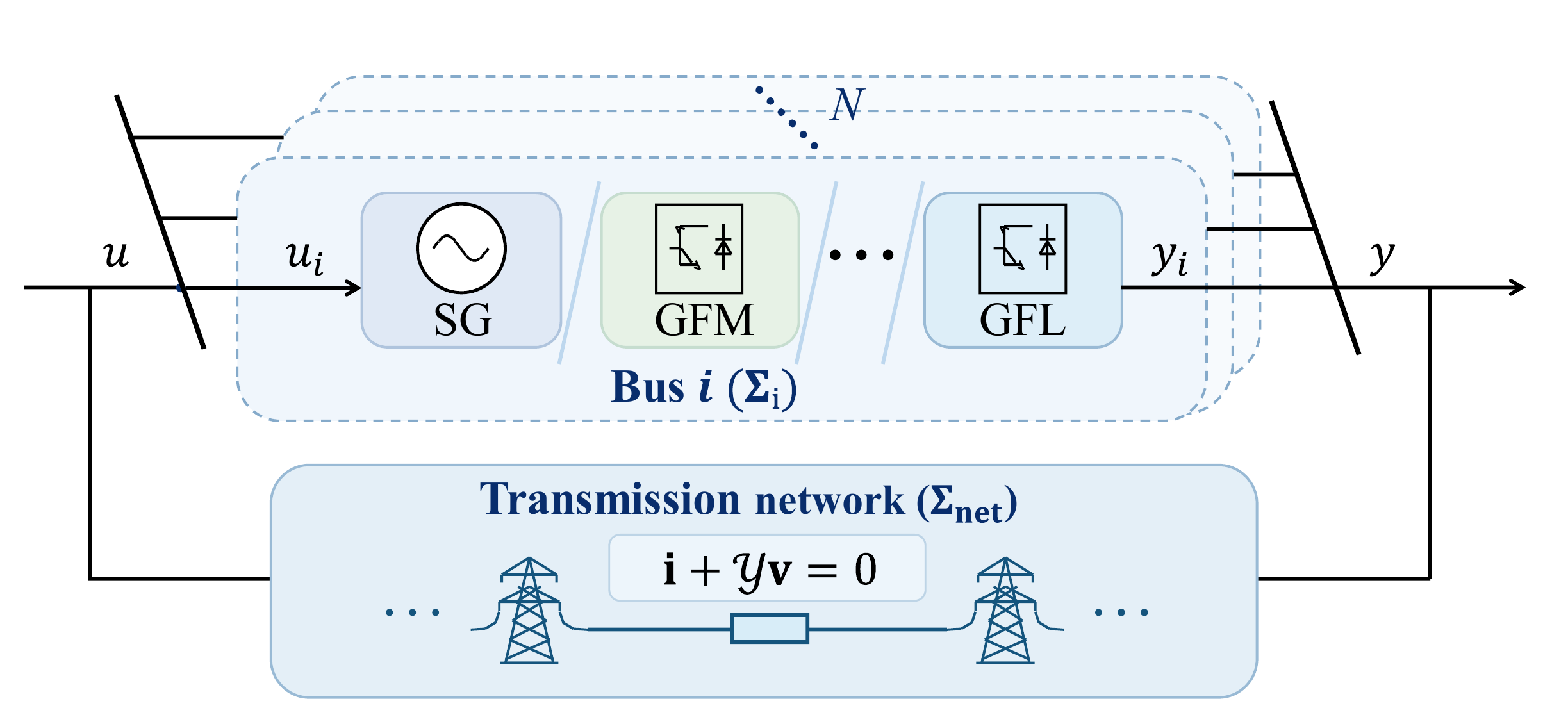}
    \caption{Feedback interconnected power system model}
    \label{fig:powersystem}
\end{figure}

Accordingly, either terminal voltage \(v_i\) or terminal current \(i_i\) may serve as the natural input variable, so dynamic and static buses may admit either causal assignment, $(u_i,y_i)=(\mathbf{v}_i,\mathbf{i}_i)$ or $(u_i,y_i)=(\mathbf{i}_i,\mathbf{v}_i)$. 
Let $\mathbf{v}=\operatorname{col}(\mathbf{v}_1,\ldots,\mathbf{v}_N)$ and $\mathbf{i}=\operatorname{col}(\mathbf{i}_1,\ldots,\mathbf{i}_N)$. With the bus-admittance matrix $Y=G+jB$ and the power flow relationship $I_Q+jI_D=Y(V_Q+jV_D)$, Kirchhoff's current law then gives
\begin{equation*}
   \mathbf{i}+\mathcal{Y}\mathbf{v}=0,\quad \mathcal Y=
\left[
\begin{array}{cc}
G_{ij} & B_{ij}\\
-B_{ij} & G_{ij}
\end{array}
\right]_{i,j=1}^{N}
\end{equation*}
where $\mathcal Y\in\mathbb R^{2N\times 2N}$
is the bus-wise real representation of $Y$. The network coupling can be represented abstractly as a static relation between the network interconnection variables, $\Sigma_{\mathrm{net}}:0=g_{\mathrm{net}}(u_{\mathrm{net}},y_{\mathrm{net}})$ where \(u_{\mathrm{net}}=y=\operatorname{col}(\{y_i\}_{i\in\mathcal{N}})\) and \(y_{\mathrm{net}}=-u=\operatorname{col}(\{u_i\}_{i\in\mathcal{N}})\) under the feedback interconnection. Combining the bus dynamics with this network relation yields the feedback-interconnected power-system model depicted in Fig.~\ref{fig:powersystem}. Collecting all internal dynamic states of the buses into $x=\operatorname{col}(\{x_i\}_{i\in\mathcal{N}_{\rm d}})$. Let $f=\operatorname{col}(\{f_i\}_{i\in\mathcal{N}_{\rm d}})$, $h=\operatorname{col}(\{h_i\}_{i\in\mathcal{N}})$, and $g(x,u):=g_{\mathrm{net}}\!\left(h(x,u),-u\right)$. The whole power system can be modeled as
\begin{equation}
        \dot{x}=f(x,u),\quad
        0=g(x,u)
    \label{eq:whole-power-system}
\end{equation}

\begin{definition}
    $(x^\star,u^\star)$ is an operating equilibrium of~\eqref{eq:whole-power-system} if $f(x^\star,u^\star)=0$, $g(x^\star,u^\star)=0$. The corresponding variable at the operating equilibrium is denoted by the superscript $\star$.
\end{definition}

For the subsequent stability analysis, deviation coordinates are introduced for all state and terminal variables, that is $\Delta x_i=x_i-x_i^\star$, $\Delta u_i=u_i-u_i^\star$ and $\Delta y_i=y_i-y_i^\star$. For notational simplicity, the prefix $\Delta$ is omitted hereafter. 

The alternative input--output orientations $(u_i,y_i)$ are useful for device-level modeling but will not be imposed on the compositional analysis developed in this paper. Instead, all buses are represented through the common physical port $p_i=\operatorname{col}(\mathbf{v}_i,\mathbf{i}_i)\in\mathbb{R}^4$, irrespective of their internal causality. Let $S_i$ select the voltage coordinates associated with bus $i$, that is $S_i := e_i^\top\otimes I_2=
\begin{bmatrix}
0_{2\times 2(i-1)} & I_2 & 0_{2\times 2(N-i)}
\end{bmatrix}
\in \mathbb{R}^{2\times 2N}$.
The physical port of each bus can consequently be expressed directly from the network voltage vector as $\mathbf{v}_i=S_i \mathbf{v}$ and $\mathbf{i}_i=-S_i\mathcal{Y}\mathbf{v}$, 
or equivalently,
\begin{equation}
    p_i=T_{p,i}\mathbf{v},\qquad
   T_{p,i}:= \begin{bmatrix} S_i;\, -S_iY \end{bmatrix} 
    \label{eq-networkrelation}
\end{equation}

 Using the common port \(p_i\), the explicit input--output orientation of each bus is no longer required in the subsequent compositional analysis. We denote the corresponding nonlinear local bus behavior compactly by
\begin{equation}
\Sigma_i:\begin{cases}
    \mathcal F_i(\dot x_i,x_i,p_i)=0,\quad &i\in\mathcal{N}_d\\
\mathcal F_i(p_i)=0,\quad &i\in\mathcal{N}_s
\end{cases}
\label{eq:bus-model-pi}
\end{equation}
where \(\mathcal F_i\) is induced by the original bus model together with the appropriate voltage--current causality.
The collection of~\eqref{eq-networkrelation} and~\eqref{eq:bus-model-pi} constitutes a structure-preserving nonlinear DAE description of the interconnected system in the common physical-port coordinates.

The resulting common-port DAE model provides a uniform basis for compositional analysis while retaining the local dynamics and causalities of individual buses. However, their dissipative characteristics may vary across frequency, whereas a static quadratic supply rate imposes a frequency-independent metric on the physical port. This motivates a common dynamic supply representation that captures frequency-dependent dissipativity while preserving the network composition structure.

\section{Unified Dynamic Supply Rate for Stability Analysis via Port Shaping}

\subsection{Unified Dynamic Supply Rate Induced by Port Shaping}\label{sec-shaping-port}
To obtain a unified supply representation for heterogeneous buses, the native physical port $p_i=\col(\mathbf v_i,\mathbf i_i)$ is augmented through common first-order shaping dynamics:
\begin{equation}
    z_i=\col(p_i,r_i),\qquad
    \dot r_i=-\omega_cr_i+\omega_cp_i
    \label{eq:first-order-lifting}
\end{equation}
where $r_i\in\mathbb{R}^{4}$ is the local shaped port and $\omega_c>0$ is the cutoff frequency. Under zero initial conditions, the corresponding Laplace-domain representation is given by
\begin{equation*}
    \widehat z_i(s)=\Psi(s)\widehat p_i(s),\quad
    \Psi(s)=
    \begin{bmatrix}
        I_4\\
        F(s)I_4
    \end{bmatrix},
    \quad
    F(s)=\frac{\omega_c}{s+\omega_c}
\end{equation*}
The resulting dynamic supply rate is parameterized as $w_i=z_i^TX_iz_i$, where $X_i=X_i^T\in\Ssym^8$ is the local supply matrix. 
The scalar shaping transfer functions $F(s)$ are shared by all buses, ensuring compatibility with the static network interconnection. 

The common first-order shaping dynamics in~\eqref{eq:first-order-lifting} possess the following properties:

\textbf{1) Network-compatible dynamic augmentation:}
To preserve the network interconnection in the augmented coordinates, a common shaped network voltage $r_{\mathbf{v}}\in\mathbb{R}^{2N}$ is introduced as $\dot r_{\mathbf{v}}=-\omega_c r_{\mathbf{v}}+\omega_c \mathbf{v}$, and the local shaped physical-port variable is $r_{i}=T_{p,i}r_{\mathbf{v}}$. The local and network dynamical ports are initialized consistently, i.e., $r_{i}(0)=T_{p,i}r_{\mathbf{v}}(0)$. Taking $q=\operatorname{col}(\mathbf{v},r_{\mathbf{v}})\in\R^{4N}$ as the common dynamic network coordinate and $T_i=\operatorname{diag}(T_{p,i},T_{p,i})\in\mathbb{R}^{8\times 4N}$ as the corresponding static port projection. According to the network relation~\eqref{eq-networkrelation}, it follows that $z_i=\operatorname{col}(T_{p,i}\mathbf{v},T_{p,i}r_{\mathbf{v}})=T_iq$.
Thus, the dynamic augmentation preserves the original network projection structure, allowing heterogeneous nodal supply rates to be composed through the common dynamic network coordinate.

\textbf{2) Complementary low- and high-frequency decomposition:}
The dynamic port~\eqref{eq:first-order-lifting} admits an equivalent representation in terms of complementary low- and high-frequency components $\tilde{z}_i=\operatorname{col}(r_i,p_i-r_i)=Hz_i$, $H=
    \begin{bmatrix}
        0,\,I_4;\,
        I_4,\,-I_4
    \end{bmatrix}$.
Accordingly, the dynamic supply rate admits the equivalent quadratic representation $w_i=z_i^TX_iz_i=\tilde{z}_i^T\widetilde{X}_i\tilde{z}_i$, with $\widetilde{X}_i=H^{-T}X_iH^{-1}$. Under zero initial conditions, the two components of $\tilde{z}_i$ satisfy $\widehat r_i=F_L(s)\widehat p_i$ and $\widehat p_i-\widehat r_i=F_H(s)\widehat p_i$ where $F_L(s)=\frac{\omega_c}{s+\omega_c}$, $F_H(s)=\frac{s}{s+\omega_c}$.
These filters satisfy \(F_L(s)+F_H(s)=1\), so the two components exactly reconstruct the original physical port, while \(|F_L(j\omega)|^2+|F_H(j\omega)|^2=1\) gives a complementary frequency-dependent magnitude partition. Denote the symmetric supply matrix as \(\tilde X_i=[\tilde X_{i,LL},\tilde X_{i,LH};\tilde X_{i,HL},\tilde X_{i,HH}]\), where \(\tilde X_{i,HL}=\tilde X_{i,LH}^{\top}\). Under zero initial conditions, the corresponding frequency-domain representation of the quadratic integral supply is characterized by
\begin{equation}
    \begin{aligned}
        &\Pi_i(j\omega)
=\Phi^H(j\omega)\tilde X_i\Phi(j\omega)\\
&=\!\frac{
\omega_c^2\tilde X_{i,LL}
\!+\!\omega^2\tilde X_{i,HH}\!+\!j\omega_c\omega
(\tilde X_{i,LH}\!-\!\tilde X_{i,LH}^{\top})}{\omega_c^2+\omega^2}
    \end{aligned}
\end{equation}
where \(\Phi(j\omega)=\operatorname{col}(F_L(j\omega)I_4,F_H(j\omega)I_4)\). This representation shows that the proposed dynamic supply rate admits distinct low- and high-frequency supply metrics, with \(\Pi_i(0)=\tilde X_{i,LL}\) and \(\lim_{\omega\to\infty}\Pi_i(j\omega)=\tilde X_{i,HH}\). 
This characterization illustrates the frequency-dependent supply-rate flexibility enabled by the proposed port shaping.

\textbf{3) Compatibility with established dissipativity supplies:}
A static QSR supply \(w_i=p_i^\top K_ip_i\), with \(K_i=K_i^\top\), is recovered by choosing $\tilde X_i=
\begin{bmatrix}
K_i,\,K_i;\,
K_i,\,K_i
\end{bmatrix}$ for which \(\Pi_i(j\omega)=K_i\) at all frequencies. Hence, static QSR supplies constitute a special case of the proposed frequency-dependent representation. Beyond static QSR supplies, the proposed representation also accommodates filtered counterparts of conventional ODP and IDP supplies. Let \(b_i=\operatorname{col}(u_i,y_i)=P_ip_i\), where \(P_i\) is the permutation matrix mapping the physical-port ordering \(p_i=\operatorname{col}(v_i,i_i)\) to the prescribed input–output ordering, and let \(E_u=[I_2\ 0]\) and \(E_y=[0\ I_2]\). Define the filtered derivatives \(d_{u,i}=\omega_cE_uP_i(p_i-r_i)\) and \(d_{y,i}=\omega_cE_yP_i(p_i-r_i)\). Under zero filter initial conditions, these correspond to \(D_{\omega_c}(s)u_i\) and \(D_{\omega_c}(s)y_i\), respectively, where \(D_{\omega_c}(s)=\omega_cs/(s+\omega_c)\). The resulting filtered ODP and IDP supply rates are $w_i^{\mathrm{ODP},f}=(u_i-\rho_i y_i)^\top d_{y,i}$, $w_i^{\mathrm{IDP},f}=d_{u,i}^\top(y_i-\nu_i u_i)$, where \(\rho_i,\nu_i\geq0\). Both are quadratic forms in the augmented port \(z_i\) and hence admit corresponding symmetric supply matrices \(X_i\). As \(\omega_c\to\infty\), \(D_{\omega_c}(j\omega)\to j\omega\) at each fixed frequency, approaching the conventional differential supply structures in the frequency-domain limit. Thus, the proposed first-order shaping unifies static QSR supplies exactly and differential-passivity supplies in filtered form within a unified, network-composable representation, enabling heterogeneous dissipativity descriptions to be jointly exploited in compositional stability analysis.


\subsection{Compositional Stability with Unified Dynamic Dissipativity}


Consider the interconnected power-system~\eqref{eq:whole-power-system}. Combining~\eqref{eq:whole-power-system} with the proposed unified dynamic port~\eqref{eq:first-order-lifting}, and defining $\chi_i=\operatorname{col}(x_i,r_i)$ ($x_i=\emptyset$ for $i\in\mathcal{N}_s$), the augmented interconnected system is expressed as
\begin{equation}
\begin{aligned}
    \widetilde{\mathcal F}_i(\dot{\chi}_i,\chi_i,p_i)=0,\quad 
    i\in\mathcal{N}
\end{aligned}
\label{eq:augmented-interconnected-system}
\end{equation}

Under the augmented dynamic port $z_i$, each bus $i\in\mathcal{N}$ is locally dissipative with respect to the quadratic supply matrix $X_i=X_i^\top$ if there exists a continuously differentiable p.s.d. storage function $V_i(\chi_i)$, satisfying
\begin{equation}
   \dot V_i(\chi_i)
    \leq z_i^\top X_i z_i
    \label{eq-nonlinear-local-dissipation}
\end{equation}
along every admissible local trajectory of the augmented bus model. For any collection of local dissipativity descriptions $\{X_i\}_{i=1}^N$, define the aggregate storage $V_{\mathrm{tot}}:=\sum_{i=1}^{N} V_i$ and the aggregate supply matrix $S(X):=\sum\nolimits_{i=1}^{N} T_i^\top X_i T_i$. Since $z_i=T_iq$, summing the local dissipation inequalities gives
\begin{equation}
    \dot V_{\mathrm{tot}}
    \leq
    \sum\nolimits_{i=1}^{N} z_i^\top X_i z_i
    =
    q^\top S(X)q
    \label{eq-nonlinear-aggregate-dissipation}
\end{equation}


\begin{definition}[$q$-detectability]
The augmented interconnected system~\eqref{eq:augmented-interconnected-system} is locally $q$-detectable at the equilibrium if there exists a neighborhood $\mathcal N$ of the equilibrium on the consistent state manifold such that every admissible trajectory of~\eqref{eq:augmented-interconnected-system} that remains in $\mathcal N$ for all $t\ge 0$ and satisfies $q(t)\equiv 0$ converges to the equilibrium.
\end{definition}


\begin{theorem}[Compositional stability criterion]\label{thm:compositional-stability}
Consider the interconnected power-system~\eqref{eq:augmented-interconnected-system} with an equilibrium at the origin. Suppose in a neighborhood of the equilibrium, the consistent states form a smooth invariant manifold \(\mathcal M\) on which the DAE induces a locally unique smooth vector field. Assume each bus admits a local dynamic dissipativity description~\eqref{eq-nonlinear-local-dissipation} and \(V_{\mathrm{tot}}\) is locally positive definite on \(\mathcal M\). Then:
\begin{enumerate}[leftmargin=1.2em,labelsep=0.4em,itemindent=0pt,listparindent=0pt]
    \item If $S(X)\preceq 0$, the equilibrium is locally Lyapunov stable.
    \item If the interconnected system~\eqref{eq:augmented-interconnected-system} is locally
    $q$-detectable and there exist a matrix $W\succ0$ and a scalar
    $\eta>0$ such that
    \begin{equation}
        M(X,\eta):=S(X)+\eta W\preceq0
        \label{eq-nonlinear-asymptotic-condition}
    \end{equation}
    then the equilibrium is locally asymptotically stable.
\end{enumerate}
\end{theorem}
\begin{proof}
According to~\eqref{eq-nonlinear-aggregate-dissipation}, $\dot V_{\mathrm{tot}}\leq0$ under $S(X)\preceq 0$. Since $V_{\mathrm{tot}}$ is locally positive definite on the consistent state manifold, it is a local Lyapunov function for the augmented interconnected system, establishing local Lyapunov stability. Under~\eqref{eq-nonlinear-asymptotic-condition}, $\dot V_{\mathrm{tot}}\leq-\eta q^\top W q\leq 0$. Choose a sufficiently small compact sublevel set of \(V_{\rm tot}\) on \(\mathcal M\) contained in the neighborhood where the above conditions hold. Since \(\dot V_{\rm tot}\le0\), this set is positively invariant. By the invariance principle, every trajectory in this set approaches the largest invariant subset of $\{\dot V_{\mathrm{tot}}=0\}$. Since $W\succ0$ and $\eta>0$, $\dot V_{\mathrm{tot}}=0$ implies $q=0$. Local $q$-detectability then implies that the only invariant limiting behavior compatible with $q(t)\equiv0$ is convergence to the equilibrium. Hence the equilibrium is locally asymptotically stable.
\end{proof}

\begin{remark}
    Unlike static QSR supplies, whose augmented representations impose identical low- and high-frequency blocks, the proposed dynamic port allows distinct supply metrics across frequency regimes. This matters in heterogeneous networks, where the local metrics enabling dissipativity complementarity at low frequencies may differ from those required at high frequencies. Static QSR must use a single frequency-independent set of local supply matrices compatible across the entire frequency range. The dynamic representation instead jointly accommodates distinct frequency-dependent metrics under a common network compatibility condition, potentially reducing conservatism in compositional stability certification.
\end{remark}

Since the added shaping dynamics are auxiliary and do not alter the physical dynamics, initializing them consistently transfers the stability conclusion to the original nonlinear system~\eqref{eq:whole-power-system}. Theorem~\ref{thm:compositional-stability} consequently reduces stability analysis to finding locally realizable matrices $X_i$ compatible through the network inequality. The next section formalizes this search.

\section{Stability Verification via Convex Dissipativity Certificates}

\subsection{Centralized Compositional Stability Verification}
For each bus \(i\), let \(\mathcal D_i\subseteq\mathbb R^{n_i+4}\) be a prescribed local validity domain of the original physical variables \((x_i,p_i)\) around the operating equilibrium. This domain is independent of the port augmentation. Let \(\mathcal V_i\) be a prescribed finite-dimensional convex cone of continuously differentiable, p.s.d. storage functions \(V_i(\chi_i)\) defined on the corresponding augmented state domain. Define the local dissipativity-certificate set $\mathcal K_i^0:=\left\{(X_i,V_i):X_i=X_i^\top,\; V_i\in\mathcal V_i,\;\eqref{eq-nonlinear-local-dissipation}\right\}$.
For a fixed \(\mathcal D_i\), both the storage-function class and the dissipation inequality are closed under nonnegative linear combinations. Hence, \(\mathcal K_i^0\) is a convex cone and is generally unbounded.

For the optimization problems developed below, we restrict the scale of the certificates by a prescribed bounded convex normalization set $\mathcal C_i$ containing a neighborhood of the origin in the certificate space, and define $\mathcal K_i:= \mathcal K_i^0\cap\mathcal C_i$.
The normalization set \(\mathcal C_i\) is chosen such that $\mathcal K_i$ is nonempty and compact. Since $\mathcal C_i$ contains a neighborhood of the origin in the certificate space, any nonzero feasible certificate in \(\mathcal K_i^0\) can be scaled into \(\mathcal K_i\). Thus, the normalization restricts the certificate magnitude while preserving the feasible certificate directions. Using the normalized local certificate sets \(\mathcal K_i\), the compositional stability condition in Theorem~\ref{thm:compositional-stability} yields the following centralized verification problem.


\begin{corollary}[Centralized compositional stability verification]
    Consider the interconnected power-system~\eqref{eq:augmented-interconnected-system} and let \(W\succ0\) be fixed. Define the centralized stability verification problem
    \begin{equation}
\begin{aligned}
 \eta^\star=&\max_{\eta,\{X_i,V_i\}}\quad \eta\\
    {\rm s.t.}\quad
    &M(X,\eta)=\eta W+\sum\nolimits_{i=1}^{N}T_i^\top X_iT_i\preceq0,\\
    &(X_i,V_i)\in\mathcal K_i,\qquad i=1,\ldots,N
\end{aligned}
\label{eq:centralized_sdp}
\end{equation}
If \(\eta^\star>0\) and the associated aggregate storage is locally positive definite, then the equilibrium is locally Lyapunov stable. In addition, if the system~\eqref{eq:augmented-interconnected-system} is locally \(q\)-detectable, then the equilibrium is locally asymptotically stable.
\label{cor:centralized_verification}
\end{corollary}

Problem~\eqref{eq:centralized_sdp} selects locally realizable supply matrices to achieve the optimal network-level matching of heterogeneous dissipativity within the structured dynamic family. The optimal value \(\eta^\star\) therefore quantifies certificate compatibility under the prescribed sets \(\mathcal K_i\), rather than a physical stability margin.

\subsection{Computational Realizations of Local Certificates}
The proposed compositional stability framework applies to both linear and nonlinear systems. Depending on the local dynamics and storage-function parameterization, the certificate search in~\eqref{eq:centralized_sdp} can be realized through different computational techniques. Representative realizations for nonlinear and linear systems are presented below.

\subsubsection{Nonlinear-system realization}
For nonlinear systems admitting polynomial models or certified polynomial approximations over the prescribed validity domains $\mathcal D_i$, the local certificate search can be formulated as Semidefinite Programming (SDP) using SOS relaxation~\cite{Parrilo2000SOS, PapachristodoulouPrajna2005SOS}. Let \(\xi_i\) collect the local variables involved in the polynomial representation, including at least \(\chi_i\) and \(p_i\), and let $\mathcal P_{\mathrm{SOS}}[\xi_i]$ denote the set of SOS polynomials in $\xi_i$. For a polynomial storage function \(V_i(\chi_i)\), a sufficient certificate for the dissipation inequality ~\eqref{eq-nonlinear-local-dissipation} is
\begin{equation}
\begin{aligned}
    z_i^\top X_i z_i-\dot V_i=\sigma_{i0}(\xi_i)+\sigma_i^\top(\xi_i)b_i(\xi_i)+\ell_i^\top(\xi_i)a_i(\xi_i)
\end{aligned}
\label{eq:SOS}
\end{equation}
where \(a_i(\xi_i)=0\) and \(b_i(\xi_i) \ge 0\) encode the retained model equalities and the prescribed domain $\mathcal{D}_i$, respectively, and $\sigma_{i0}\in\mathcal P_{\mathrm{SOS}}[\xi_i]$, $\sigma_i\in \mathcal P_{\mathrm{SOS}}[\xi_i]^{\mathrm{dim}(b_i)}$, $\ell_i\in\mathbb{R}[\xi_i]^{\mathrm{dim}(a_i)}$. Representation~\eqref{eq:SOS} guarantees nonnegativity of the dissipation residual over the admissible local set.
With fixed validity domains and prescribed polynomial degrees, the resulting certificate search is a finite-dimensional convex SDP.

\subsubsection{Linear-system specialization}
Linearizing the local bus model around the prescribed operating equilibrium and incorporating the filter dynamics~\eqref{eq:first-order-lifting} yields $\tilde{E}_i\dot\chi_i=\tilde{A}_i\chi_i+\tilde{B}_ip_i$, $z_i=\tilde{C}_i\chi_i+\tilde{D}_ip_i$.
The linear dynamics impose \(G_i\operatorname{col}(\chi_i,p_i,\dot\chi_i)=0\), where \(G_i=[-\widetilde A_i\;\;-\widetilde B_i\;\;\widetilde E_i]\). Let $N_i$ form a basis of $\ker G_i$. For the quadratic storage $V_i=\chi_i^\top Q_i\chi_i$, $Q_i\succeq0$, the local dissipation inequality~\eqref{eq-nonlinear-local-dissipation} is ensured by
\begin{equation}
    N_i^\top\left\{\begin{bmatrix}
0&0&Q_i\\
0&0&0\\
Q_i&0&0
\end{bmatrix}-\begin{bmatrix}
\tilde C_i\\\tilde D_i\\0
\end{bmatrix} X_i\begin{bmatrix}
\tilde C_i\\\tilde D_i\\0
\end{bmatrix}^\top\right\}N_i\preceq0
    \label{eq:local_dissipativity_lmi}
\end{equation}

\section{Physics-Informed Distributed Verification}

\subsection{Network-Critical Dissipation Direction}

To enable a distributed implementation of the centralized stability verification problem~\eqref{eq:centralized_sdp}, we first examine its dual structure. We introduce a dual matrix \(\Lambda\succeq0\) associated with the network compatibility constraint. The Lagrangian is
\begin{align*}
    \mathcal L&=\eta-\left\langle\Lambda,\,\eta W+\sum\nolimits_{i=1}^N T_i^\top X_iT_i\right\rangle\\
    &=\eta\left(1-\langle W,\Lambda\rangle\right)-\sum\nolimits_{i=1}^N \langle T_i \Lambda T_i^\top,X_i\rangle
\end{align*}
Maximization over the unrestricted $\eta$ requires $\ip{W}{\Lambda}=1$. Define the compact dual set \(\Delta_W:=\{\Lambda\succeq0:\langle W,\Lambda\rangle=1\}\). For $\Lambda\in\Delta_W$, the local minimum support value is
\begin{equation}
    \phi_i(\Lambda)
    =
    \min_{(X_i,V_i)\in\mathcal K_i}
    \ip{\Gamma_i(\Lambda)}{X_i},\quad \Gamma_i(\Lambda)=T_i\Lambda T_i^\top
    \label{eq:phi}
\end{equation}
Since the local certificate sets \(\mathcal K_i\) are nonempty, compact, and convex, and the network constraint admits a strictly feasible point for sufficiently negative \(\eta\), strong duality gives
\begin{equation}
    \eta^\star
    =
    \min_{\Lambda\in\Delta_W}
    \left[-\sum\nolimits_{i=1}^{N}\phi_i(\Lambda)\right]
    \label{eq:exact-dual}
\end{equation}

The optimal dual matrix is supported on the network-critical generalized eigenspace.

\begin{proposition}[Network-Critical Dual Weighting]
\label{prop:critical_dual}

Let \(\{X_i^\star,V_i^\star\}_{i=1}^{N}\), \(\eta^\star\), and \(\Lambda^\star\in\Delta_W\) be primal and dual optimal solutions of~\eqref{eq:centralized_sdp} and~\eqref{eq:exact-dual}, respectively. Define $S^\star=S(X^\star)=\sum_{i=1}^{N} T_i^\top X^\star_i T_i$ and $M^\star=S^\star+\eta^\star W$. Then
\begin{equation}
M^\star\Lambda^\star=0,\quad\eta^\star
=
-\lambda_{\max}
\left(
W^{-1/2}S^\star W^{-1/2}
\right)
\label{eq:network-critical}
\end{equation}
Thus, $\operatorname{range}\Lambda^\star\subseteq\ker M^\star$, and every nonzero vector in \(\operatorname{range}(\Lambda^\star)\) is a generalized eigenvector of \((S^\star,W)\) associated with \(\lambda_{\max}(S^\star,W)\).
\end{proposition}

\begin{proof}
    Complementary slackness gives $\langle \Lambda^\star,-M^\star\rangle=0$. Since \(\Lambda^\star\succeq0\) and \(-M^\star\succeq0\), it follows that \(M^\star\Lambda^\star=0\). For fixed \(X^\star\), the largest feasible \(\eta\) satisfying \(S^\star+\eta W\preceq0\) is $-\lambda_{\max}\left(W^{-1/2}S^\star W^{-1/2}\right)$, which yields~\eqref{eq:network-critical}.
\end{proof}
If the maximum generalized eigenvalue is simple and \(v\) spans \(\ker M^\star\), the normalization \(\langle W,\Lambda^\star\rangle=1\) gives $\Lambda^\star=\frac{vv^\top}{v^\top Wv}$. For a multiple active eigenvalue, \(\Lambda^\star\) instead provides a positive-semidefinite weighting over the critical subspace and need not have full support on that subspace. For any nonzero network coordinate $q$, the generalized Rayleigh quotient $(q^\top S^\star q)/(q^\top W q)$ measures the normalized aggregate supply along the network direction \(q\). At bus $i$, in the simple-eigenvalue case, the corresponding local weighting is $\Gamma_i(\Lambda^\star)=T_i\Lambda^\star T_i^\top={(T_i v)(T_i v)^\top}/{v^\top Wv}$.
Thus, \(\Gamma_i(\Lambda^\star)\) transfers the global network-critical dissipation information to the local augmented port and weights local certificate improvement according to its relevance to the current network bottleneck. This matrix-valued weighting provides the coordination signal used in the distributed certificate-exchange procedure developed next.

\subsection{Distributed Network-Critical Margin Exchange (NCME)}

We now exploit the dual characterization above to solve the centralized verification problem through iterative exchange between the network and individual buses. 
Algorithm~\ref{alg:ncme} summarizes the procedure.
\begin{enumerate}[
    label=\arabic*),
    leftmargin=*,
    labelindent=0pt,
    itemindent=0pt,
    align=left,
    labelsep=0.5em]
    \item \textbf{Initialization}. Each bus initializes a finite feasible certificate library \(\mathcal A_i^0\subseteq\mathcal K_i\) containing the zero certificate. Unless otherwise specified, \(\mathcal A_i^0=\{(0,0)\}\). To obtain an informative initial atom, the network selects $\Lambda_{\rm iso}^0=\frac{W^{-1}}{4N},\quad\langle W,\Lambda_{\rm iso}^0\rangle=1$ and broadcasts \(\Gamma_i^0=T_i\Lambda_{\rm iso}^0T_i^\top\). 
    \item \textbf{Restricted master}. At iteration \(k\), let $\mathcal A_i^{k-1}=\{(X_i^{(j)},V_i^{(j)})\}_{j=0}^{m_i^{k-1}}$
    where \(m_i^{k-1}\) is the current size of the local certificate library. Define $Y_i^{(j)}=T_i^\top X_i^{(j)}T_i$. The network solves the restricted master problem
\begin{equation}
\begin{aligned}
\eta_k=&\max_{\eta,\{\alpha_{ij}\}}\quad  \eta\\[-1mm]
\mathrm{s.t.}\quad
& \eta W+\sum\nolimits_{i=1}^{N}\sum\nolimits_{j=0}^{m_i^{k-1}}
\alpha_{ij}Y_i^{(j)}\preceq0,\\[-1mm]
& \sum\nolimits_{j=0}^{m_i^{k-1}}\alpha_{ij}=1,\quad
\alpha_{ij}\ge0,\ \forall i,j
\end{aligned}
\label{eq:RMP}
\end{equation}
    Whenever the certificates satisfy the storage condition of Corollary~\ref{cor:centralized_verification} with \(\eta_k>0\), local Lyapunov stability is certified without requiring convergence of the optimization. Local \(q\)-detectability may be checked separately when asymptotic stability is desired.
    \item \textbf{Network-critical pricing}. Let \(\Lambda^k\) be an optimal dual matrix of~\eqref{eq:RMP}. The network broadcasts $\Gamma_i^k=T_i\Lambda^kT_i^\top$ and each bus independently solves the local pricing problem~\eqref{eq:phi} and obtains $\phi_i(\Lambda^k)$ and $(X_i^{k,+},V_i^{k,+})\in\arg\min_{(X_i,V_i)\in\mathcal K_i}\left\langle\Gamma_i^k,X_i\right\rangle$. 
    Exact pricing gives the current dual upper bound, and its best-so-far value is retained as
    \begin{equation}
        \eta_{U}^k=-\sum\nolimits_i\phi_i(\Lambda^k)\ge\eta^\star,\quad \bar\eta_k=\min\nolimits_{j\le k}\eta_{U}^j
        \label{eq:dual-bound}
    \end{equation}
    The newly priced certificate is added to the library as \(\mathcal A_i^{k}=\mathcal A_i^{k-1}\cup\{(X_i^{k,+},V_i^{k,+})\}\).
\item \textbf{Termination}. The procedure terminates when either a storage-qualified positive-margin certificate is obtained or $\bar\eta_k-\eta_k\le\epsilon_{\rm gap}$. In the latter case, the centralized margin is resolved within tolerance, and the absence of a strict certificate is reported as \emph{not certified}.
\end{enumerate}

\begin{algorithm}[t]
\caption{Network-Critical Margin Exchange (NCME)}
\label{alg:ncme}
\begin{algorithmic}[1]
\REQUIRE Local sets $\mathcal K_i$, maps $T_i$, $W\succ0$,
tolerances 
$\epsilon_{\rm gap}$.

\STATE Initialize
$\mathcal A_i^0\subseteq\mathcal K_i$, $(0,0)\in\mathcal A_i^0$, $|\mathcal A_i^0|<\infty$, $\bar\eta_0=+\infty$.
\STATE Set
$\Lambda_{\rm iso}^0=W^{-1}/(4N)$; broadcast
$\Gamma_i^0=T_i\Lambda_{\rm iso}^0T_i^\top$,
and augment each $\mathcal A_i^0$ by one parallel pricing solve.

\FOR{$k=1,2,\ldots$}

\STATE Solve~\eqref{eq:RMP} over $\operatorname{conv}\mathcal A_i^{k-1}$,
obtain $\eta_k$, $\alpha_{ij}^k$,
and $\Lambda^k$.

\IF{$\eta_k$ > 0 and the aggregate storage is locally positive definite}
    \STATE \textbf{return} certified stable.
\ENDIF

\STATE Broadcast
$\Gamma_i^k=T_i\Lambda^kT_i^\top$; each bus solve~\eqref{eq:phi} in parallel and returns
$(X_i^{k,+},V_i^{k,+})$ and the local pricing value.

\STATE Compute $\eta_{U}^k$ and $\bar\eta_k$ through~\eqref{eq:dual-bound}.

\IF{$\bar\eta_k-\eta_k\le\epsilon_{\rm gap}$}
    \STATE return margin resolved; stability not certified
\ENDIF

\STATE Add $(X_i^{k,+},V_i^{k,+})$ to its certificate library.

\ENDFOR
\end{algorithmic}
\end{algorithm}

\begin{proposition}[Certified bounds and convergence]
    Suppose that each \(\mathcal K_i\) is nonempty, compact, and convex, the restricted-master and local pricing problems are solved exactly. Then $\eta_k\le\eta^\star\le\bar\eta_k$ for all \(k\), with \(\{\eta_k\}\) nondecreasing and \(\{\bar\eta_k\}\) nonincreasing. Moreover, for the iteration continued without early termination, $\eta_k\uparrow\eta^\star$, $\bar\eta_k\downarrow\eta^\star$, and hence $\bar\eta_k-\eta_k\rightarrow0$.
    \label{prop:convergence}
\end{proposition}
\begin{proof}

The bounds and monotonicity follow directly from restriction, weak duality, and nested certificate libraries. Since $\{\eta_k\}$ is nondecreasing and bounded above by $\eta^\star$, let $\eta_k\to\ell\le\eta^\star$.
Define
$h(\Lambda,X):=-\sum_i
\langle T_i\Lambda T_i^\top,X_i\rangle$
and
$f(\Lambda):=-\sum_i\phi_i(\Lambda)$.
Let $f_k^R$ denote the restricted dual objectiveat iteration $k$, so that $\eta_k=f_k^R(\Lambda^k)$. By compactness of $\prod_i\mathcal K_i$, $f$ and all $f_k^R$ are uniformly Lipschitz continuous on $\Delta_W$.
Select a subsequence
$\Lambda^{k_j}\to\widehat\Lambda$.
Uniform Lipschitz continuity gives $f_{k_j}^R(\widehat\Lambda)
-f_{k_j}^R(\Lambda^{k_j})\to 0$. Since $f_{k_j}^R(\Lambda^{k_j})
=\eta_{k_j}\to\ell$ and
$f_k^R(\widehat\Lambda)$ is nondecreasing,
it follows that
$f_k^R(\widehat\Lambda)\to\ell$. Exact pricing at $k_j$ generates
$X^{k_j,+}\in\prod_i\mathcal K_i$
satisfying
$h(\Lambda^{k_j},X^{k_j,+})
=f(\Lambda^{k_j})$.
By uniform continuity of $h$ on the compact
feasible sets, $h(\widehat\Lambda,X^{k_j,+})
\to f(\widehat\Lambda)$. Since the newly generated pricing atoms are retained in the nested certificate libraries, $f_{k_j+1}^R(\widehat\Lambda)
\ge h(\widehat\Lambda,X^{k_j,+})$. Passing to the limit yields
$\ell\ge f(\widehat\Lambda)
\ge\eta^\star$.
Together with $\ell\le\eta^\star$,
this proves $\eta_k\uparrow\eta^\star$. Moreover, along the same subsequence,
$\eta_U^{k_j}=f(\Lambda^{k_j})
\to f(\widehat\Lambda)=\eta^\star$.
Since $\bar\eta_k=\min_{t\le k}\eta_U^t$
is nonincreasing and satisfies
$\bar\eta_k\ge\eta^\star$, it follows that
$\bar\eta_k\downarrow\eta^\star$.
Consequently, $\bar\eta_k-\eta_k\to0$.
\end{proof}


\section{Case Studies}


\subsection{Heterogeneous Four-Bus Benchmark}

\begin{figure}
    \centering
    \includegraphics[width=0.85\linewidth]{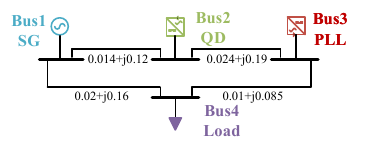}
    \caption{Topology of the heterogeneous four-bus system.}
    \label{fig:hete4bus}
\end{figure}

\begin{figure}
    \centering
    \includegraphics[width=1\linewidth]{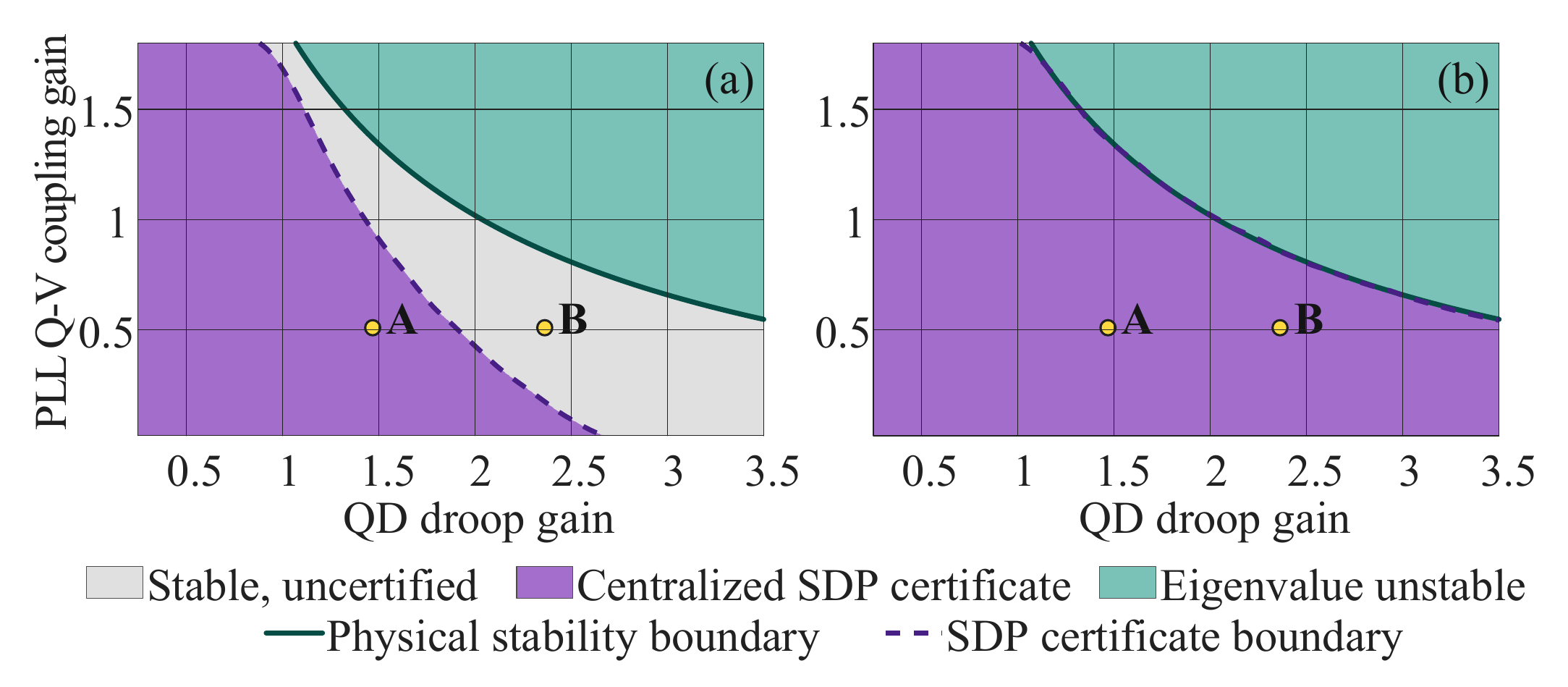}
    \caption{Centralized stability certification of the four-bus benchmark using (a) the static port and (b) the dynamic port.}
    \label{fig:fourbus4d8d}
\end{figure}

We first consider the heterogeneous four-bus benchmark shown in Fig.~\ref{fig:hete4bus}, consisting of an SG, a QD-GFM converter, a PLL-GFL converter, and a static ZIP load. To assess the conservatism reduction in stability analysis enabled by the dynamic port, the normalized QD gain and PLL reactive-voltage coupling gain are varied over a two-dimensional parameter grid. Finite generalized eigenvalues of the complete structure-preserving DAE are used as the stability reference.
Fig.~\ref{fig:fourbus4d8d} compares centralized certificates obtained with the native static port \(p_i\) and the filtered dynamic port \(z_i\). The static formulation certifies $63.9\%$ of the eigenvalue-stable operating points, whereas the dynamic formulation certifies $100\%$, with no unstable points certified. Full certification coverage is retained for \(\omega_c\in[0.1,10]\) p.u., indicating low sensitivity to the cutoff-frequency selection. 

\begin{figure}
    \centering
    \includegraphics[width=1\linewidth]{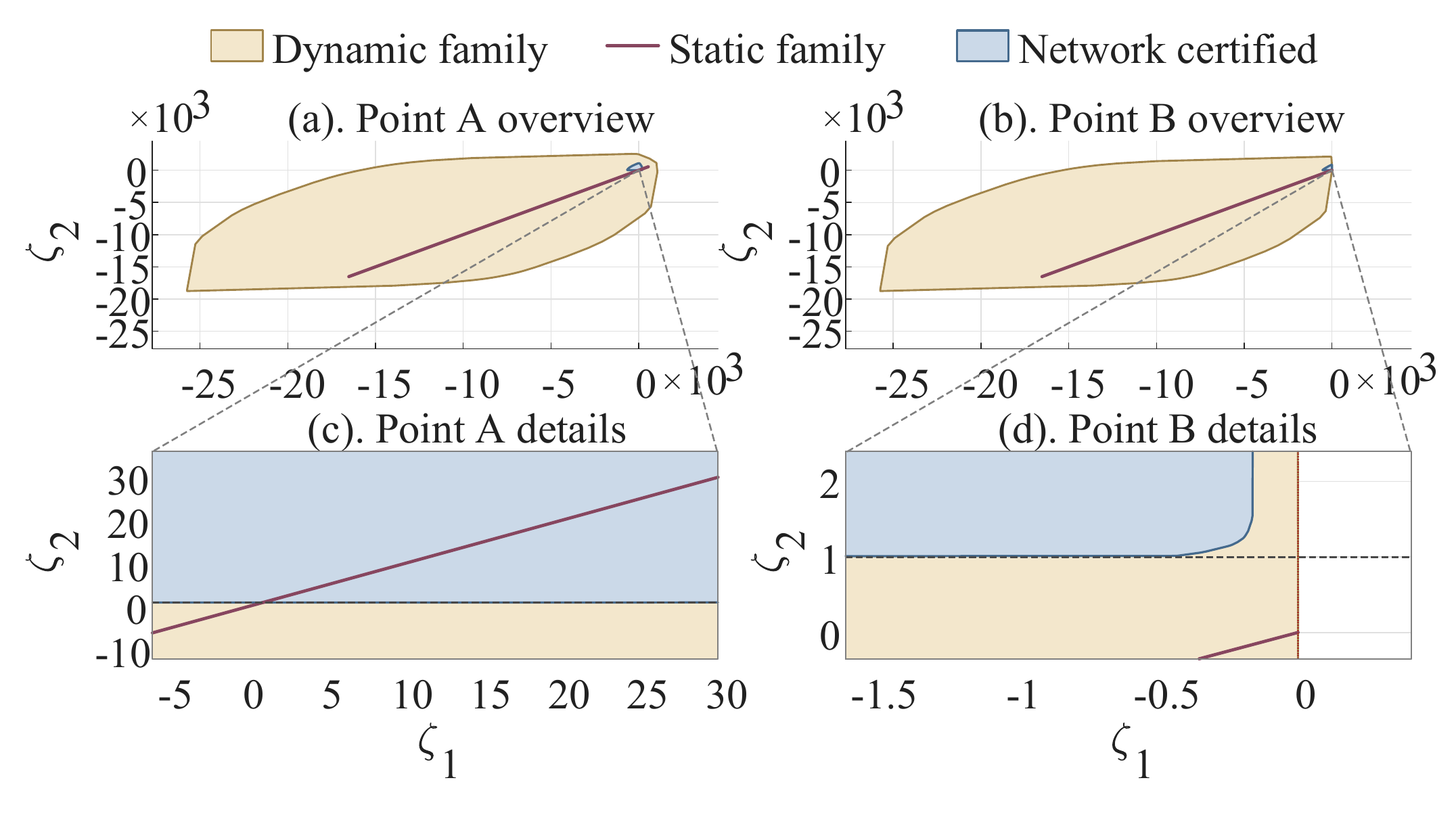}
    \caption{Additional dissipativity freedom and network certification at points A and B. (a), (b) Global projections; (c), (d) enlarged views.}
    \label{fig:dynamic_port_mech}
\end{figure}

To explain the certification advantage in Fig.~\ref{fig:fourbus4d8d}, we examine two representative stable points. Point A is certifiable with both ports, and point B is certifiable only with the dynamic port. Their static- and dynamic-port certificate families are projected onto a common plane with coordinates $s_1=-\langle\Lambda^p,S(X)\rangle$ and $s_2=\sum_i(\mu_i^p)^\top\mathcal R_i(X_i)$, representing network dissipative support and a selected direction of frequency-dependent supply-rate variation, respectively. Here, \(\Lambda^p\in\mathbb S^{4N}\) and \(\mu_i^p\in\mathbb R^{16}\) are fixed common projection weights, $\mathcal R_i(X_i)=\operatorname{col}\big(\operatorname{ut}(\widetilde X_{i,LL}-\widetilde X_{i,HH}),\operatorname{sut}(\widetilde X_{i,LH}-\widetilde X_{i,LH}^{\top})\big)$, where \(L\) and \(H\) denote the low- and high-frequency channels, and \(\operatorname{ut}\) and \(\operatorname{sut}\) extract the upper-triangular entries with and without the diagonal in column-wise order, respectively. The projected regions are computed by certificate feasibility tests, with the network-certified subset obtained by additionally imposing the complete certification conditions. The displayed coordinates are \(\zeta_1=(s_1-s_2)/\eta_0\) and \(\zeta_2=s_1/\eta_0\), with \(\eta_0=10^{-5}\). According to Section~\ref{sec-shaping-port}, \(\mathcal R_i(X_i)=0\) characterizes frequency-independent supply metrics. The static family is therefore confined to \(\zeta_1=\zeta_2\), whereas off-diagonal points reflect frequency-dependent variation in the induced supply metrics. As illustrated in Fig.~\ref{fig:dynamic_port_mech}, static supplies suffice for network certification at A but not at B, where the dynamic family admits locally realizable and network-compatible certificates outside the static family. This illustrates how the supply-rate flexibility enabled by dynamic port shaping facilitates dissipativity matching beyond static supply descriptions.

\begin{figure}
    \centering
    \includegraphics[width=1\linewidth]{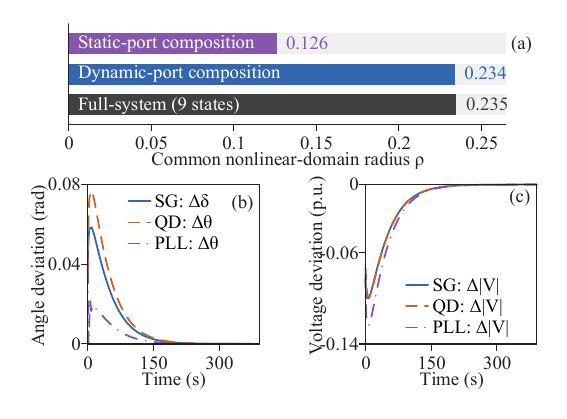}
    \caption{Nonlinear stability certification of the heterogeneous four-bus system. (a) Certified validity-domain radii. (b) Angle and (c) voltage responses within the dynamic-port certificate validity domain.}
    \label{fig:fourbus_nonlinear}
\end{figure}


We further examine the benefits of dynamic port shaping for nonlinear stability certification. Three formulations are compared: the conventional static-port formulation, the proposed dynamic-port formulation, and a noncompositional full-system benchmark. To ensure a fair comparison, a common validity domain \(\mathcal D_i(\rho)\) is prescribed by bounding the normalized deviations of the independent physical variables \((x_i,u_i)\) from their operating-equilibrium values \(\|\operatorname{col}(x_i,u_i)\|_2\leq\rho\). Here, the port variables \(p_i\) are determined by the original device equations rather than treated as independent variables, so this constraint induces a common validity domain in the original physical variables \((x_i,p_i)\) for all three formulations. Under identical SOS relaxation settings, the maximum certifiable radii are searched for the two-port formulations and the full-system benchmark. As shown in Fig.~\ref{fig:fourbus_nonlinear}(a), the proposed dynamic-port formulation enlarges the certified validity-domain radius from 0.126 to 0.234, closely approaching the full-system result of 0.235. The corresponding nonlinear responses in Figs.~\ref{fig:fourbus_nonlinear}(b) and (c) illustrate convergence from an initial condition on the boundary of the certified validity domain at \(\rho=0.234\). These results show that the proposed dynamic-port formulation not only improves certification coverage across operating points but also substantially enlarges the validity domains of nonlinear compositional certificates.


To assess distributed certification efficiency, NCME is compared with ADMM at three stable operating points representing deep-interior, intermediate-margin, and near-boundary conditions, termed Deep, Middle, and Near. Both use identical local certificate sets and scaling bounds, a common target of $1\%$ of the centralized dynamic-port optimal margin, and the same normalized certificate audit. Each method terminates upon finding the first passing certificate or reaching the prescribed time limit. Table~\ref{tab:ncme_admm} reports the corresponding timing costs across ten runs per point. The reported computational costs are averaged over the independent runs. Overall, NCME achieves substantially lower coordination and computation costs than ADMM across the tested operating points.


\begin{table}[t]
\centering
\caption{NCME and ADMM Certification Performance
on the Nonlinear Four-Bus Benchmark}
\label{tab:ncme_admm}
\begin{tabular}{llccc}
\hline
Point & Method & Succ. & Rounds & Wall-clock time (s) \\
\hline
Deep
& NCME & 10/10 & 13 & 1.88 \\
& ADMM & 10/10 & 64 & 6.197 \\
\hline
Middle
& NCME & 10/10 & 15 & 2.01 \\
& ADMM & 10/10 & 86 & 7.81 \\
\hline
Near
& NCME & 10/10 & 24 & 2.45 \\
& ADMM & 0/10 & 2188 & 200.64 \\
\hline
\end{tabular}
\end{table}

\subsection{Modified IEEE 118-Bus System}

\begin{figure}
    \centering
    \includegraphics[width=1\linewidth]{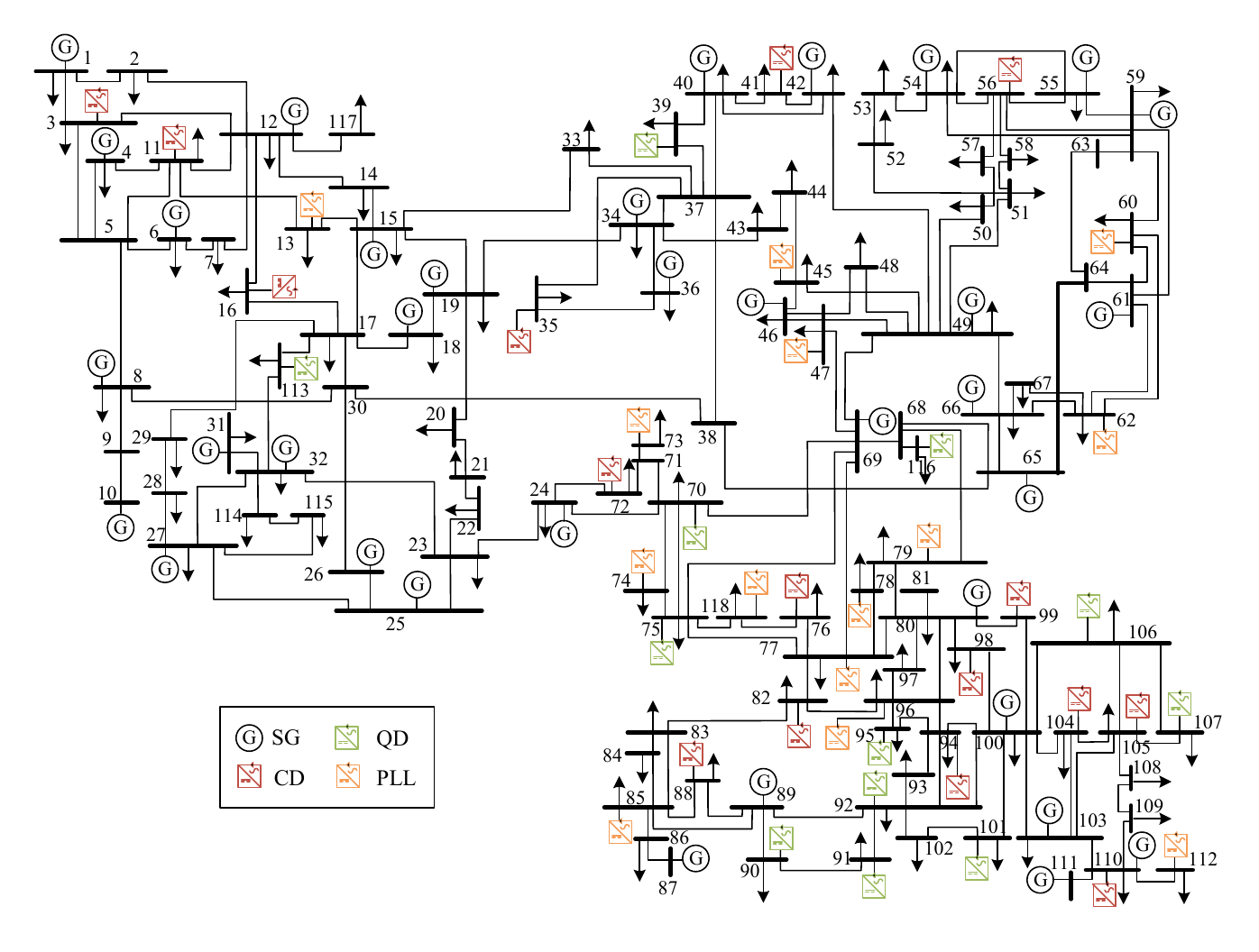}
    \caption{The structure of the modified IEEE 118-Bus benchmark.}
    \label{fig:ieee118model}
\end{figure}

\begin{figure}
    \centering
    \includegraphics[width=1\linewidth]{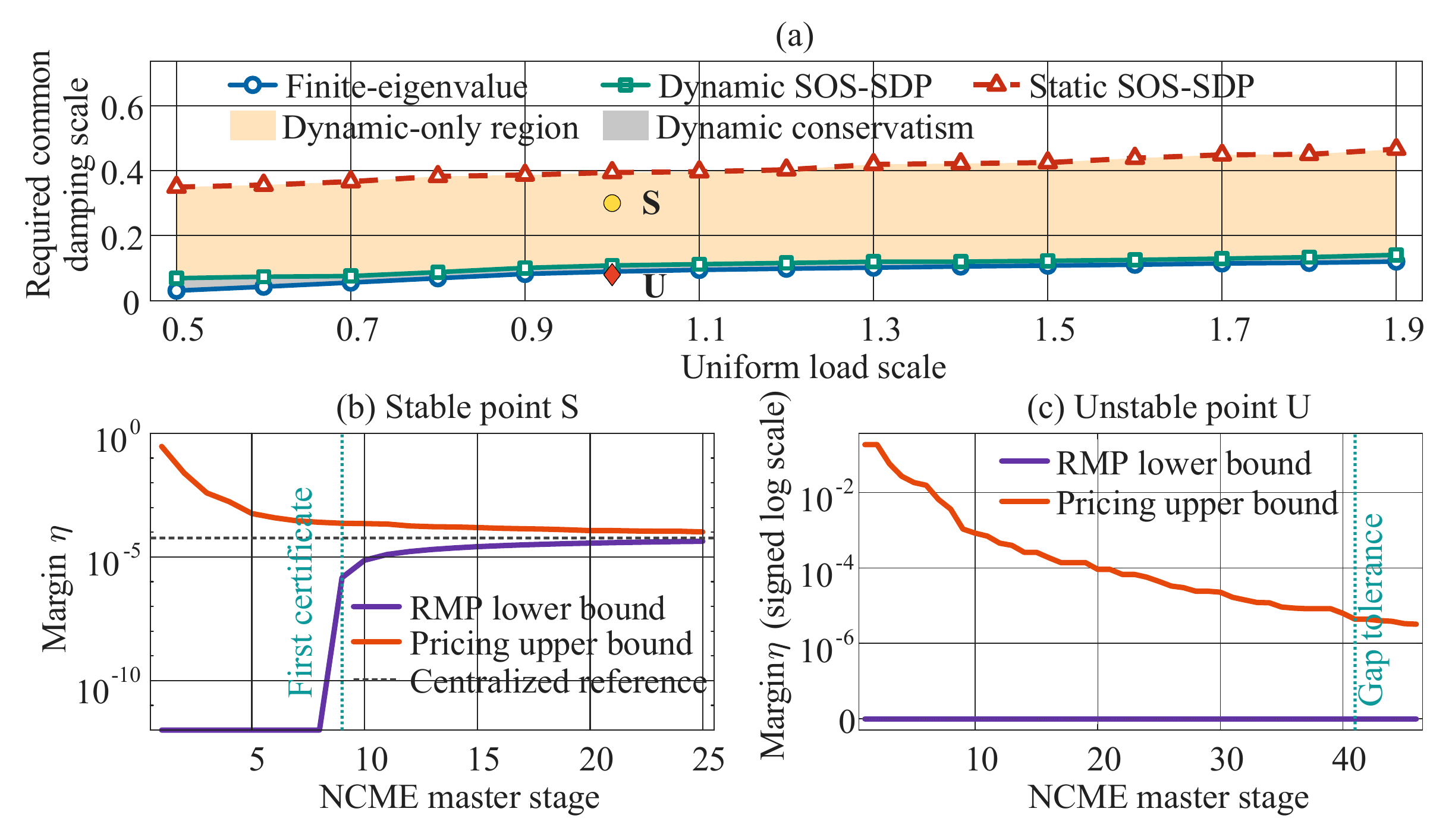}
    \caption{Compositional stability certification of the modified IEEE 118-bus system. (a) Stability and certification boundaries. NCME bound evolution at (b) stable point S and (c) unstable point U.}
    \label{fig:ieee118_boundary}
\end{figure}

To evaluate the proposed framework on a large-scale heterogeneous system, a modified IEEE 118-bus system is constructed on the original benchmark~\cite{5491276}. As illustrated in Fig.~\ref{fig:ieee118model}, the system contains 34 synchronous generators (SGs), 15 CD-GFM converters, 14 QD-GFM converters, and 14 PLL-GFL converters, together with static ZIP loads and algebraic junction buses.
To span a broad operating range, the uniform load scale \(\lambda_L\) is varied from 0.5 to 1.9, while a common damping scale \(\kappa_d\) is applied to the nominal damping coefficients of all dynamic devices. For each \(\lambda_L\), the critical damping scales are determined by finite-eigenvalue analysis and centralized SOS-based certification with the static and dynamic ports, respectively. As shown in Fig.~\ref{fig:ieee118_boundary}(a), the dynamic-port formulation requires substantially less damping for certification. It brings the certification boundary much closer to the finite-eigenvalue stability limit, demonstrating reduced conservatism in large-scale stability analysis.

We further examine the distributed NCME procedure at two representative operating points S and U, marked in Fig.~\ref{fig:ieee118_boundary}(a). As shown in Fig.~\ref{fig:ieee118_boundary}(b), at the certifiable stable point S, the NCME bounds progressively approach the centralized optimum, with the first valid certificate obtained at stage \(k=9\), well before full bound convergence. At the finite-eigenvalue-unstable point U, Fig.~\ref{fig:ieee118_boundary}(c) shows that the lower bound remains at numerical zero, while the upper bound decreases until the prescribed gap tolerance is reached at \(k=41\), without obtaining a valid certificate. These results demonstrate that NCME agrees with centralized certification and illustrate its two termination mechanisms in large-scale systems.

\begin{figure}[t]
    \centering
    \includegraphics[width=1\linewidth]{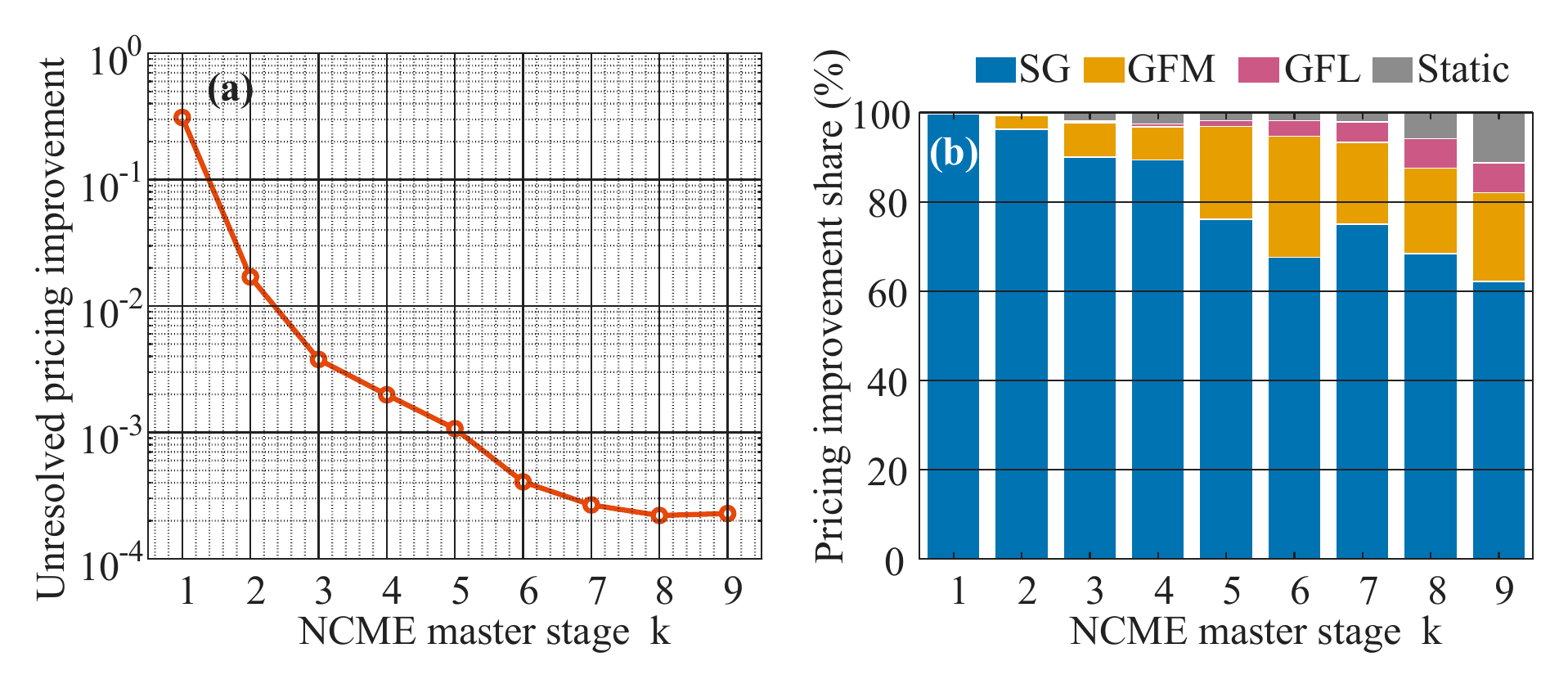}
    \caption{NCME convergence and certificate-enrichment process in the modified IEEE 118-Bus system. (a) Aggregate unresolved pricing improvement. (b) Device-class composition of the pricing improvement.}
    \label{fig:ieee118ncme}
\end{figure}

\begin{table}[t]
\centering
\caption{Computational Performance Comparison of ADMM and NCME
on the IEEE 118-Bus System at Operating Point S}
\label{tab:admm_ncme_118}
\begin{tabular}{lcccc}
\hline
Method & Success & Total Time (s) & Updates & Local SOS Calls \\
\hline
ADMM & 3/3 & 583.33 & 487 & 57,348 \\
NCME & 3/3 & 99.32 & 8 & 1,180 \\
\hline
\end{tabular}
\end{table}

To expose the underlying exchange mechanism, define the restricted support $\phi_{i,\mathcal A}^k=\min_{(X_i,V_i)\in\mathcal{A}_i^{k-1}}\ip{\Gamma_i^k}{X_i^j}$, the pricing improvement $\delta_i^k:=\phi_{i,\mathcal A}^k-\phi_i(\Lambda^k)\ge0$, and the aggregate unresolved pricing improvement $R_k:=\sum_{i=1}^{N}\delta_i^k$. $R_k$ measures the total improvement still available from the full local certificate sets relative to the current libraries under the critical weighting \(\Lambda_k\). As shown in Fig.~\ref{fig:ieee118ncme}(a), \(R_k\) decreases by more than three orders of magnitude before certification, confirming that successive NCME exchanges progressively reduce the remaining local certificate improvements not represented by the restricted libraries. Fig.~\ref{fig:ieee118ncme}(b) further decomposes \(R_k\) by device class. The unresolved pricing improvement is initially almost entirely associated with SG buses, but progressively redistributes toward GFM, PLL-GFL, and static/algebraic buses as the certificate libraries are enriched. Hence, NCME does not repeatedly target a predetermined subset of devices. Instead, the network-critical weighting adaptively steers certificate generation toward the local sets that offer the largest remaining pricing improvement.

The computational benefits of this network-critical coordination mechanism are further evaluated by comparing NCME with ADMM at operating point S under identical SOS certificate settings and certification criteria. As shown in Table~\ref{tab:admm_ncme_118}, both methods achieve successful certification in all three runs, whereas NCME requires only 8 coordination updates and 1,180 local SOS solves, compared with 487 and 57,348 for ADMM, respectively. These results demonstrate the substantial computational efficiency gains NCME offers for large-scale nonlinear stability certification.

Further benchmark parameters and additional implementation details of the nonlinear certification procedures are provided in Appendix B.

\section{Conclusion}

This paper develops a compositional stability-analysis framework for heterogeneous power systems that combines a unified dynamic dissipativity representation with scalable distributed stability certification. A common first-order augmentation is deployed to shape the voltage–current ports, enabling flexible local supply rates with complementary frequency-dependent information. It represents heterogeneous buses within a unified, network-composable framework while substantially reducing the conservatism of stability certification. The resulting compositional certification problem was solved through dual-guided column generation. Its matrix dual identifies the network-critical dissipation direction, which provides a physically meaningful coordination signal for parallel local certificate generation in the NCME method. The lower–upper bound mechanism enables distributed certification and stability verification before full optimization convergence. Case studies on nonlinear heterogeneous power-system benchmarks demonstrate the improved certification capability enabled by dynamic port shaping, as well as the computational scalability and reduced coordination overhead of NCME.
Future work will focus on identifying system-specific dynamic supply rates that best capture available dissipativity, together with data-driven extensions for online stability assessment.

\bibliographystyle{IEEEtran}
\bibliography{reference.bib}

\appendices

\renewcommand{\thedefinition}{\thesection.\arabic{definition}}
\renewcommand{\theassumption}{\thesection.\arabic{assumption}}
\renewcommand{\thelemma}{\thesection.\arabic{lemma}}
\renewcommand{\theremark}{\thesection.\arabic{remark}}
\renewcommand{\thetheorem}{\thesection.\arabic{theorem}}
\renewcommand{\thecondition}{\thesection.\arabic{condition}}
\renewcommand{\theproposition}{\thesection.\arabic{proposition}}

\setcounter{definition}{0}
\setcounter{assumption}{0}
\setcounter{lemma}{0}
\setcounter{remark}{0}
\setcounter{theorem}{0}
\setcounter{condition}{0}
\setcounter{proposition}{0}

\counterwithin{equation}{section}


\section{Typical Heterogeneous Node Models and Dissipativity}
\label{app:bus_models}\label{app:conventional_dissipativity}
\label{app:augmented_dissipativity}\label{sec-app-unified-port-representation}
\counterwithin{equation}{section}\setcounter{equation}{0}

In the main text's $D$--$Q$ ordering, let $v=(V_D,V_Q)^\top$, $i=(I_D,I_Q)^\top$, $J=\left[\begin{smallmatrix}0&1\\-1&0\end{smallmatrix}\right]$, $V=\|v\|_2$, and $s(\theta)=(\sin\theta,\cos\theta)^\top$. Injection powers are $P=-v^\top i$ and $Q=-v^\top Ji$. Write $\Delta a=a-a^\star$, $\bar p=\operatorname{col}(v,i)$, and $p=\Delta\bar p$. The models below expand Table~\ref{tab:heterogeneous_nodes}; polar descriptions require positive magnitudes and a continuous local angle branch. With $z=\operatorname{col}(p,r)$ defined by~\eqref{eq:first-order-lifting}, a static QSR supply has the exact representation
\begin{equation}
 X^{\rm QSR}=\operatorname{diag}(K,0_4)
 =H^\top\begin{bmatrix}K&K\\K&K\end{bmatrix}H
 \label{eq:app_qsr_augmented}
\end{equation}
The two matrices act on $z$ and $\widetilde z=Hz$, respectively. For a fixed linear port map $b=\operatorname{col}(u_c,y_c)=Lp$, define $E_u=[I_2\ 0]$, $E_y=[0\ I_2]$, $U=[E_uL\ 0]$, $Y=[E_yL\ 0]$, $D_u=\omega_c[E_uL\ {-E_uL}]$, and $D_y=\omega_c[E_yL\ {-E_yL}]$. The filtered differential supplies are represented by
\begin{equation}
\begin{aligned}
 \mathcal X_{\rm ODP}(L,\rho_i)&=\operatorname{sym}((U-\rho_iY)^\top D_y)\\
 \mathcal X_{\rm IDP}(L,\nu_i)&=\operatorname{sym}(D_u^\top(Y-\nu_iU))
\end{aligned}
\label{eq:app_differential_templates}
\end{equation}
where $\operatorname{sym}A=(A+A^\top)/2$. Thus $z^\top\mathcal X_{\rm ODP}z=(u_c-\rho_i y_c)^\top d_y$ and $z^\top\mathcal X_{\rm IDP}z=d_u^\top(y_c-\nu_i u_c)$, with $d_u=D_uz$, $d_y=D_yz$.

The power/polar maps used below are nonlinear. For $b=\phi(\bar p)-\phi(\bar p^\star)$, write $b=Lp+e_\phi(p)$, where $L=D\phi(\bar p^\star)$ and $e_\phi(p)=O(\|p\|_2^2)$. Equation~\eqref{eq:app_differential_templates} represents the filtered differential supplies associated with the tangent map $Lp$. The corresponding dynamic-port dissipativity is assessed through a storage $V_i(x_i,r_i)$ satisfying~\eqref{eq-nonlinear-local-dissipation}. Finite-domain certification additionally retains the nonlinear coordinate relations and their remainders.

\subsection{Synchronous Generators}

\subsubsection{Typical node model}
A one-axis generator behind transient reactance $x'_d>0$ has internal voltage $e=E s(\delta)$ and voltage-input/current-output relation $i=J(e-v)/x'_d$. At the internal port define $P_e=-e^\top i$ and $Q_e=-e^\top Ji$. A representative governor and excitation model is
\begin{equation*}
\begin{aligned}
 \dot\delta&=\omega\\
 M\dot\omega&=-D_g\omega-K_I\Delta\delta-\Delta P_e\\
 T_d\dot E&=-k_E\Delta E-d_x\Delta(Q_e/E)
\end{aligned}
\end{equation*}
Here $d_x=x_d-x'_d>0$, $D_g$ includes proportional governor damping, and $k_E$ includes excitation feedback. The field equation is the equilibrium-centered form of $T_d\dot E=a-k_EE-d_xQ_e/E$. The lossless transient reactance gives $P_e=P$ and $Q_e=Q+x'_d\|i\|_2^2$.

\subsubsection{Conventional dissipativity}
For $M,D_g,T_d,K_I,k_E>0$, use the power/polar port $u_c=(-\Delta P_e,-\Delta(Q_e/E))^\top$, $y_c=(\Delta\delta,\Delta E)^\top$, and storage
\begin{equation*}
 W=\tfrac12M\omega^2+\tfrac12K_I(\Delta\delta)^2
                    +\frac{k_E}{2d_x}(\Delta E)^2
\end{equation*}
Multiplying the swing equation by $\omega$ and the field equation by $\dot E/d_x$ yields
\begin{equation*}
 \dot W=u_c^\top\dot y_c-D_g\omega^2-\frac{T_d}{d_x}\dot E^2
\end{equation*}
Consequently, $W_{\rho_i}=W-\rho_i\|y_c\|_2^2/2$ satisfies $\dot W_{\rho_i}\leq(u_c-\rho_i y_c)^\top\dot y_c$. It is locally positive definite for $0\leq\rho_i<\min\{K_I,k_E/d_x\}$, giving the ODP structure in Table~\ref{tab:heterogeneous_nodes}.

\subsubsection{Dynamic-port compatibility}
The internal voltage is $e=v-x'_dJi=T\bar p$, with $T=[I_2\ {-x'_dJ}]$. For this map, and for the terminal-port special case $T=[I_2\ 0]$, put $I_i=[0\ I_2]$ and define
\begin{equation}
\begin{aligned}
 h_\theta&=[e_2^\star\ {-e_1^\star}]/(E^\star)^2\qquad
 h_E=(e^\star)^\top/E^\star\\
 G_P&=(i^\star)^\top T+(e^\star)^\top I_i\\
 G_Q&=\frac{(Ji^\star)^\top T+(e^\star)^\top JI_i}{E^\star}
       +\frac{Q_e^\star(e^\star)^\top T}{(E^\star)^3}\\
 L_g(T)&=\operatorname{col}(G_P,G_Q,h_\theta T,h_ET)
\end{aligned}
\label{eq:app_polar_jacobian}
\end{equation}
This is the Jacobian of $\operatorname{col}(-P_e,-Q_e/E,\arg\underline e,E)$. The generator's tangent filtered-ODP matrix is $X_{\rm SG}^{\rm ODP,f}=\mathcal X_{\rm ODP}(L_g(T),\rho_i)$; the full nonlinear map supplies the coordinate relations for finite-domain compatibility.

\subsection{Virtual Synchronous Generators}

\subsubsection{Typical node model}
A representative VSG uses virtual inertia and damping with reactive-power droop. Its current-input/voltage-output realization is $v=V s(\theta)$, with
\begin{equation*}
\begin{aligned}
 \dot\theta&=\omega\\
 M_v\dot\omega&=-D_v\omega-K_\theta\Delta\theta-\Delta P\\
 \tau_v\dot V&=-\Delta V-D_q\Delta Q
\end{aligned}
\end{equation*}
All displayed controller coefficients are positive, and the voltage loop is considered near an equilibrium with $V^\star>0$.

\subsubsection{Conventional dissipativity}
For $c_v=1/D_q+Q^\star/V^\star>0$, define
\begin{equation}
 \Psi(V)=c_v\big[V-V^\star-V^\star\ln(V/V^\star)\big]
 \label{eq:app_droop_potential}
\end{equation}
Using $\Psi'(V)=c_v\Delta V/V$ and the voltage equation, the storage $W=M_v\omega^2/2+K_\theta(\Delta\theta)^2/2+\Psi(V)$ satisfies, for $u_c=(-\Delta P,-\Delta(Q/V))^\top$ and $y_c=(\Delta\theta,\Delta V)^\top$,
\begin{equation*}
 \dot W=u_c^\top\dot y_c-D_v\omega^2-\frac{\tau_v}{D_qV}\dot V^2
\end{equation*}
The shifted storage $W_{\rho_i}=W-\rho_i\|y_c\|_2^2/2$ is locally positive definite for $0\leq\rho_i<\min\{K_\theta,c_v/V^\star\}$, establishing a representative ODP supply.

\subsubsection{Dynamic-port compatibility}
Here $e=v$, so use $T=[I_2\ 0]$ in~\eqref{eq:app_polar_jacobian}. The tangent filtered matrix is $X_{\rm VSG}^{\rm ODP,f}=\mathcal X_{\rm ODP}(L_g(T),\rho_i)$, with the nonlinear power/polar map retained when constructing a finite-domain augmented storage.

\subsection{Conventional-Droop Grid-Forming Converters}

\subsubsection{Typical node model}
A conventional-droop converter generates $v=V s(\theta)$ from incoming current $i$. A representative outer-loop model is
\begin{equation*}
 \tau_\theta\dot\theta=-\Delta\theta-D_p\Delta P\qquad
 \tau_V\dot V=-\Delta V-D_q\Delta Q
\end{equation*}
with positive time constants and droop coefficients.

\subsubsection{Conventional dissipativity}
Use the same power/polar port as for VSG and the potential~\eqref{eq:app_droop_potential}. For $W=(\Delta\theta)^2/(2D_p)+\Psi(V)$, multiplication by $\dot\theta/D_p$ and $\dot V/(D_qV)$ gives
\begin{equation*}
 \dot W=u_c^\top\dot y_c-\frac{\tau_\theta}{D_p}\dot\theta^2
                           -\frac{\tau_V}{D_qV}\dot V^2
\end{equation*}
The ODP storage $W_{\rho_i}=W-\rho_i\|y_c\|_2^2/2$ is locally positive definite for $0\leq\rho_i<\min\{1/D_p,c_v/V^\star\}$. This is the conventional-droop structure represented in Table~\ref{tab:heterogeneous_nodes}.

\subsubsection{Dynamic-port compatibility}
For $T=[I_2\ 0]$, the physical-port tangent supply is $X_{\rm CD}^{\rm ODP,f}=\mathcal X_{\rm ODP}(L_g(T),\rho_i)$. The dependence on droop coefficients enters the admissible index and storage; the coordinate map is the common terminal power/polar map.

\subsection{Quadratic-Droop Grid-Forming Converters}

\subsubsection{Typical node model}
A quadratic-droop converter also has current input and output $v=V s(\theta)$, with representative dynamics~\cite{SimpsonPorco2017QuadraticDroop,Yang2020}
\begin{equation*}
\begin{aligned}
 \tau_1\dot\theta&=-\Delta\theta-D_1\Delta P\\
 \tau_2\dot V&=-D_2Q-V(V-u_\star)\qquad
 u_\star=V^\star+D_2Q^\star/V^\star
\end{aligned}
\end{equation*}
All time constants and droop coefficients are positive.

\subsubsection{Conventional dissipativity}
Dividing the voltage equation by $D_2V$ gives $\tau_2\dot V/(D_2V)=-\Delta(Q/V)-\Delta V/D_2$. Thus, for the same power/polar port and $W=(\Delta\theta)^2/(2D_1)+(\Delta V)^2/(2D_2)$,
\begin{equation*}
 \dot W=u_c^\top\dot y_c-\frac{\tau_1}{D_1}\dot\theta^2
                           -\frac{\tau_2}{D_2V}\dot V^2
\end{equation*}
The shifted storage establishes ODP for $0\leq\rho_i<\min\{1/D_1,1/D_2\}$. In these normalized-reactive-power coordinates, its voltage storage is quadratic.

\subsubsection{Dynamic-port compatibility}
The coordinate Jacobian is again $L_g([I_2\ 0])$, yielding $X_{\rm QD}^{\rm ODP,f}=\mathcal X_{\rm ODP}(L_g([I_2\ 0]),\rho_i)$. Its admissible indices differ from conventional droop, while both share the same physical-port transformation.

\subsection{PLL-Synchronized Grid-Following Converters}

\subsubsection{Typical node model}
A representative PLL-based GFL follows the terminal phasor $v=V s(\theta_v)$ and controls its injected powers:
\begin{equation*}
\begin{aligned}
 \dot\xi&=V\sin(\theta_v-\delta)\\
 \dot\delta&=K_PV\sin(\theta_v-\delta)+K_I\xi\\
 P&=P^\star-K_\omega\dot\delta-K_a\Delta\delta\\
 Q&=Q_r-K_{V1}V-K_{V2}(V-V^\star)V
\end{aligned}
\end{equation*}
Here $\delta^\star=\theta_v^\star$, $\xi^\star=0$, and $Q_r=Q^\star+K_{V1}V^\star$ ensures equilibrium consistency. The voltage-input/current-output realization follows from
\begin{equation*}
 i=-\frac{1}{V^2}\begin{bmatrix}P&-Q\\Q&P\end{bmatrix}v
\end{equation*}

\subsubsection{Conventional dissipativity}
The IDP port is $u_c=(\Delta\theta_v,\Delta V)^\top$, $y_c=(-\Delta P,-\Delta(Q/V))^\top$. Linearizing at the synchronized equilibrium with $x_c=(\Delta\xi,\Delta\delta)^\top$ gives $\dot x_c=A_cx_c+B_cu_c$, $y_c=C_cx_c+D_cu_c$, where
\begin{equation*}
\begin{aligned}
 A_c&=\begin{bmatrix}0&-V^\star\\K_I&-K_PV^\star\end{bmatrix}\qquad
 B_c=\begin{bmatrix}V^\star&0\\K_PV^\star&0\end{bmatrix}\\
 C_c&=\begin{bmatrix}K_\omega K_I&K_a-K_\omega K_PV^\star\\0&0\end{bmatrix}\\
 D_c&=\operatorname{diag}\left(K_\omega K_PV^\star,
                         Q_r/(V^\star)^2+K_{V2}\right)
\end{aligned}
\end{equation*}

Let $W=W(x_c,u_c)$ and regard $a=\dot u_c$ as the driving input of the prolonged realization. With $\zeta=\operatorname{col}(x_c,u_c)$, this realization is $\dot\zeta=A_e\zeta+B_ea$, where $A_e=\left[\begin{smallmatrix}A_c&B_c\\0&0\end{smallmatrix}\right]$ and $B_e=\operatorname{col}(0,I_2)$. A sufficient quadratic-storage IDP condition is

\begin{equation*}
\begin{aligned}
 Q_e&\succ0\qquad Q_eA_e+A_e^\top Q_e\preceq0\\
 Q_eB_e&=\begin{bmatrix}C_c^\top\\(D_c-\nu_iI_2)^\top\end{bmatrix}
\end{aligned}
\end{equation*}
For any feasible $Q_e$, the input-dependent storage $W(x_c,u_c)=\zeta^\top Q_e\zeta/2$ satisfies
\begin{equation*}
\begin{aligned}
 \dot W&=\tfrac12\zeta^\top(Q_eA_e+A_e^\top Q_e)\zeta
                  +\dot u_c^\top B_e^\top Q_e\zeta\\
 &\leq\dot u_c^\top(y_c-\nu_i u_c)
\end{aligned}
\end{equation*}
Thus, feasibility with $\nu_i\geq0$ establishes the representative linearized IDP description in Table~\ref{tab:heterogeneous_nodes}, using a storage defined on $(x_c,u_c)$.

\subsubsection{Dynamic-port compatibility}
Exchanging the input and output blocks of~\eqref{eq:app_polar_jacobian} gives $L_{\rm PLL}=\left[\begin{smallmatrix}0&I_2\\I_2&0\end{smallmatrix}\right]L_g([I_2\ 0])$.
Equation~\eqref{eq:app_differential_templates} gives $X_{\rm PLL}^{\rm IDP,f}=\mathcal X_{\rm IDP}(L_{\rm PLL},\nu_i)$, representing the filtered tangent IDP supply in the common voltage--current augmented port. The corresponding dynamic-port dissipativity is assessed separately through a storage $V_i(x_i,r_i)$ satisfying~\eqref{eq-nonlinear-local-dissipation}, independently of the input-dependent storage used for the conventional IDP description.

\subsection{ZIP Loads}

\subsubsection{Typical node model}
A static ZIP load has voltage input and current output. For $s=V/V_0$,
\begin{equation*}
\begin{aligned}
 P_L(V)&=P_{L0}(a_{PZ}s^2+a_{PI}s+a_{PP})\\
 Q_L(V)&=Q_{L0}(a_{QZ}s^2+a_{QI}s+a_{QP})\\
 i_L(v)&=\frac{1}{V^2}\begin{bmatrix}P_L&-Q_L\\Q_L&P_L\end{bmatrix}v
\end{aligned}
\end{equation*}
Each coefficient triple sums to one; $P_L,Q_L$ denote signed demands, opposite to network injections.

\subsubsection{Conventional dissipativity}
Let $A_L(v)=\partial i_L/\partial v$ on a convex voltage domain with positive magnitude. The identity $\Delta i_L=\int_0^1A_L(v^\star+t\Delta v)\,dt\,\Delta v$ shows that $\operatorname{sym}A_L(v)\succeq mI_2$ implies
\begin{equation*}
 \Delta v^\top\Delta i_L\geq m\|\Delta v\|_2^2\qquad
 K_{\rm ZIP}=\begin{bmatrix}-mI_2&I_2/2\\I_2/2&0\end{bmatrix}
\end{equation*}
Thus $p^\top K_{\rm ZIP}p\geq0$ is a static QSR certificate with zero storage. The lower bound $m$ may be negative for constant-power consumption.

\subsubsection{Dynamic-port compatibility}
The exact embedding is $X_{\rm ZIP}=\operatorname{diag}(K_{\rm ZIP},0_4)$ by~\eqref{eq:app_qsr_augmented}, with the same zero storage and physical validity domain.

\subsection{Constant-Impedance Loads}

\subsubsection{Typical node model}
A constant-impedance load obeys $i_L=Y_Lv$, $Y_L=G_LI_2+B_LJ$, with voltage input and current output.

\subsubsection{Conventional dissipativity}
Since $J^\top=-J$, $\Delta v^\top\Delta i_L=G_L\|\Delta v\|_2^2$. For $G_L\geq0$ this is a passive static QSR relation, with zero storage and $K_{\rm P}=\left[\begin{smallmatrix}0&I_2/2\\I_2/2&0\end{smallmatrix}\right]$.

\subsubsection{Dynamic-port compatibility}
The same supply is recovered exactly by $X_{\rm Z}=\operatorname{diag}(K_{\rm P},0_4)$ and zero storage.

\subsection{Intermediate and Junction Buses}

\subsubsection{Typical node model}
A junction without local injection satisfies $i=0$, so its voltage-input/current-output relation is algebraic and $\Delta i=0$.

\subsubsection{Conventional dissipativity}
The cross supply vanishes: $p^\top K_{\rm P}p=\Delta v^\top\Delta i=0$. Hence the junction has a zero-storage static QSR description.

\subsubsection{Dynamic-port compatibility}
Its exact physical-port embedding is $X_{\rm J}=\operatorname{diag}(K_{\rm P},0_4)$, whose supply vanishes on the junction graph.

\subsection{Constant-Voltage Sources}

\subsubsection{Typical node model}
An ideal regulated source prescribes $v=v^\star$ independently of incoming current. Its native orientation is current input and voltage output, with $\Delta v=0$.

\subsubsection{Conventional dissipativity}
Again $p^\top K_{\rm P}p=0$, now because the voltage deviation vanishes. This is a static QSR relation with zero storage.

\subsubsection{Dynamic-port compatibility}
Choosing $X_{\rm V}=\operatorname{diag}(K_{\rm P},0_4)$ recovers this supply exactly. Thus the three algebraic classes share a cross-supply matrix while imposing different physical port graphs.

\section{Benchmark Parameters and Nonlinear Certification Details}
The four-bus SG uses $M=0.1607$, $D=0.8$, $T_d=6.56$ s, $x_d=0.295$, $x'_d=0.17$, $K_P=0$, $K_I=m_{\rm SG}\sigma$, and $K_E=(x_d-x'_d)\sigma-1$, where $\sigma=2.67692969129$ at unit safety factor. QD uses $\tau_1=0.8$ s, $\tau_2=8$ s and $D_1=D_2=k_{\rm QD}/\sigma$. The radius experiment fixes $m_{\rm SG}=1$, $k_{\rm QD}=1.46875$, and PLL bandwidth parameter $b=0.30$. Its PLL has $k_p=2b$, $k_i=18b^2$, $k_a=0.30$, $\tau_i=0.16$ s, $K_{v,d}=0.25$, and $K_{q,v}=0.5095$. Both ZIP coefficient triples are $(0.20,0.30,0.50)$, with $P_{L0}=1.15\times1.75$, $Q_{L0}=1.15\times0.55$, and $V_0=1$. Equilibria and internal references are reconstructed from AC power flow.

The IEEE nominal SG has $R_s=0.002$, $X_s=0.25$, $M=5$, $D=8$, $K_p=1$, $K_a=1.5$, $T_E=1$ s, $K_Q=0.25$, and $K_V=0.50$. CD/QD have $R_s=0.025$, $X_s=0.18$, $\tau_\theta=0.25$ s, $\tau_E=0.35$ s, $K_\theta=1$, $D_p=0.65$, $D_q=0.55$, $K_{v\theta}=K_{\theta V}=0.08$, and nominal $K_{\rm quad}=1$ for QD. PLL nominal values are $k_p=2$, $k_i=18$, $k_a=0.35$, $\tau_i=0.08$ s, and $K_{v,d}=K_{q,v}=0.15$. IEEE ZIP triples are $(0.25,0.20,0.55)$ for active and $(0.30,0.20,0.50)$ for reactive demand, with $V_0=V^\star$. These are templates: bus heterogeneity $h$, damping scale $d$, and time scale $t$ multiply $M$ by $h$, each of $D,K_a,K_\theta,K_{\rm quad}$ by $hd$, and each of $T_E,\tau_\theta,\tau_E$ by $t/h$. Operating-point gain scales are applied separately. PLL gains scale by its own $h$, its restoring gain by $hd$, and $\tau_i$ by $1/h$.

The radius experiment uses local deviations $w_i$ of device states and native electrical inputs, with $\|w_i\|_2\leq\rho$ in the common per-unit convention (angles in radians). All three methods use identical prescribed certificate validity domains and device equations for comparison. Both nonlinear implementations specialize~\eqref{eq:SOS} with quadratic storage, substituting explicit dynamics and retaining algebraic graph relations where needed.

For the performance experiments, let $F_i(w_i)$ collect state derivatives and physical port components in local deviation coordinates. Write $\Delta F_i(w_i):=F_i(w_i)-F_i(0)=J_iw_i+e_i(w_i)$, with $J_i=DF_i(0)$. Componentwise second-order Taylor remainder bounds $|e_{ij}(w_i)|\leq c_{ij}\|w_i\|_2^2$, $c_{ij}\geq0$, are propagated using domain-wide derivative bounds for elementary nonlinear functions. On each prescribed ball $\|w_i\|_2\leq R_i$, set $b_{ij}:=R_ic_{ij}$ to obtain $|e_{ij}(w_i)|\leq b_{ij}\|w_i\|_2$. For $j\in\mathcal J_i:=\{j:b_{ij}>0\}$, write $e_{ij}=b_{ij}\epsilon_{ij}$ and $q_{ij}:=\|w_i\|_2^2-\epsilon_{ij}^2\geq0$; zero-bound components have $e_{ij}=0$ and need no auxiliary variable. The quadratic residual $g_i$ is certified by requiring $g_i-\sum_{j\in\mathcal J_i}\mu_{ij}q_{ij}$ to be SOS, with $\mu_{ij}\geq0$, implemented as a Gram-matrix LMI. This sufficient robust envelope condition relaxes correlations among the exact nonlinear remainder components.

The radius experiment instead jointly searches storage, supply, and polynomial SOS multipliers on the device graphs. Sine and cosine are expanded to degree 11 with the same angle variable and rigorous Taylor tails. Positive denominators are cleared (SG: factor $E^2$; ZIP: $V^4$); their positive lower bounds are checked on each domain. For PLL and ZIP, voltage magnitude is lifted by $V^2=v^\top v$, $V>0$, retaining its algebraic correlation rather than independently bounding a square root. Truncated scalar residual degrees are $(12,10,10,6)$ for (SG,QD,PLL,ZIP); discarded polynomial tails and interval coefficient errors are bounded and subtracted as a multiple of the squared local norm. Sparse SOS bases have maximum degrees $(6,5,5,3)$, with ball multipliers of maximum degrees $(10,8,8,4)$ and free equality multipliers for lifted relations. The nine-state centralized baseline searches a full quadratic Lyapunov matrix using the same original-function graph bounds; it imposes no dissipativity port structure. 

The radius search and nonlinear IEEE studies use $\omega_c=0.5$ s$^{-1}$. In the radius search, local storage traces are bounded by 20, $-20I\preceq X\preceq20I$, and the centralized storage trace by 60, with $V_i=\chi_i^\top Q_i\chi_i$. Global positive definiteness and strict network/derivative inequalities are checked separately. Reported certificates pass original-function and numerical residual checks. Search failures mean no certificate was found within the stated finite class and settings.

\end{document}